\documentclass[12pt, leqno]{article}
\usepackage{amsmath, amsthm, enumitem}
\usepackage{amsfonts}
\usepackage{amssymb}
\usepackage[usenames,dvipsnames]{color}
\usepackage{graphicx}
\usepackage{bbm}
\usepackage{hyperref}
\usepackage{bm}
\usepackage{listings}
\usepackage{color,calc}
\usepackage{subcaption}
\usepackage{cite}

\newtheorem{theorem}{Theorem}

\newtheorem{rem}{Remark}
\renewcommand{\d}{{\rm d}}

\newcommand{\Fdag}{F^\dagger}
\newcommand{\Sdag}{S^\dagger}
\newcommand{\Tdag}{T^\dagger}
\newcommand{\phidag}{\varphi^\dagger}

\makeatletter
\def\namedlabel#1#2{\begingroup
    #2%
    \def\@currentlabel{#2}%
    \phantomsection\label{#1}\endgroup
}

\title{Limit theorems for the neutron transport equation}

\author{Eric Dumonteil\thanks{CEA, Paris Saclay. E-mail: \texttt{eric.dumonteil@cea.fr}}, \ Emma Horton\thanks{
Department of Statistics, 
University of Warwick, Coventry, CV4 7AL, UK. E-mail: \texttt{\{emma.horton\}, \{andreas.kyprianou\}@warwick.ac.uk}} ,   
\ Andreas E. Kyprianou$^\dagger$
\ and
Andrea Zoia\thanks{CEA, Paris Saclay. E-mail: \texttt{andrea.zoia@cea.fr}}
 }

\begin{document}
\maketitle
\begin{abstract}
Over the last decade, ingenuous developments in Monte Carlo methods have enabled the unbiased estimation of adjoint-weighted reactor parameters expressed as bilinear forms, such as kinetics parameters and sensitivity coefficients. A prominent example is the Iterated Fission Probability method, which relies on the simulation of the fission chains descending from an ancestor neutron: the neutron population at an asymptotic fission generation yields an estimate of the importance function (and hence of the adjoint fundamental eigenmode) at the phase-space coordinates of the ancestor neutron. In this paper we first establish rigorous results concerning the moments of the asymptotic neutron population stemming from a single initial particle, with special focus on the average and the variance. Then, we propose a simple benchmark configuration where exact solutions are derived for these moments, which can be used for the verification of new functionalities of production Monte Carlo codes involving the Iterated Fission Probability method.

\smallskip


\end{abstract}

\section{Introduction}

Monte Carlo simulation is the gold standard method for radiation transport applications, since it involves a minimal number of approximations. In particular, the phase space does not need to be discretized and all the particle-nuclei interaction physics contained in the nuclear data libraries can be used \cite{LuxKoblinger}. Thanks to these features, particle-transport codes based on the Monte Carlo method allow one to establish reference solutions against which those produced by faster but approximate deterministic solvers (which rely on the discretization of the phase space) can be benchmarked for accuracy.

The stochastic version of the standard power iteration algorithm is the workhorse of Monte Carlo codes for $k$-eigenvalue problems: a collection of neutrons is followed over a sufficiently large number of fission generations, and statistics are recorded on the successive generations once the population has settled into a stationary state. At equilibrium, the ratio between the statistical weights of the neutrons at two consecutive fission generations converges to the fundamental eigenvalue $k_0$, and the neutron flux within a generation correspondingly converges to the fundamental $k$-eigenmode, $\varphi_0$ \cite{LuxKoblinger}. Over the last decade, a major breakthrough based on the rediscovery of the Iterated Fission Probability (IFP) method has enabled the use of Monte Carlo to compute adjoint-weighted quantities \cite{nauchi_adjoint_2010, kiedrowski_adjoint_2011}. The key idea is that the adjoint fundamental eigenmode of a $k$-eigenvalue calculation is proportional to the neutron importance, which can be estimated by tracking the descendants of an ancestor neutron over a sufficiently large number of power iteration generations and collecting the statistical weights of the surviving neutrons. This technique has been implemented in several production Monte Carlo codes, enabling unbiased estimates for the fundamental adjoint flux \cite{igor, tantillo} or adjoint-weighted parameters (bilinear forms) such as kinetics parameters or perturbations and sensitivities to nuclear data \cite{truchet, kiedrowski_adjoint_2013, terranova_perturbation_2018, burke_2018_0, monk_app, kiedrowski_review_sensitivity}.

When developing new algorithms and functionalities in production Monte Carlo codes such as {\sc MCNP6}\textsuperscript{\textregistered} \cite{mcnp}, MONK\textsuperscript{\textregistered} \cite{monk} or {\sc TRIPOLI-4}\textsuperscript{\textregistered} \cite{brun-tripoli}, which involve several hundred thousands of lines of code, it is of utmost importance to rely on exact solutions whenever possible for verification purposes. Several such sets of analytical solutions have been established for regular (forward) transport problems, and in particular $k$-eigenvalue calculations \cite{ganapol, sood, griesheimer}. For adjoint eigenvalue problems and related adjoint-weighted parameters, comparatively fewer exact solutions have been derived. A prominent example is the two-group, infinite medium model benchmark \cite{kiedrowski_benchmark}, which has been successfully used to verify recent developments of Monte Carlo codes \cite{truchet, kiedrowski_adjoint_2011, burke_2018_0}. In view of these considerations, in this work we set out to establish benchmark solutions for adjoint $k$-eigenvalue problems in spatially finite media, which can usefully complement those of the infinite medium model.

Since IFP methods are generally concerned with the asymptotic behaviour of a stochastic neutron population, in this paper we provide general statements about the moments of the neutron population as a function of fission generations. More precisely, we provide exact asymptotics for the moments (of any order) of the fission generation populations for the sub-, super- and critical cases. In particular, this implies that we can obtain precise asymptotics for the average and variance of the neutron population. We further introduce a benchmark configuration where exact results can be derived. For this purpose, we resort to the `rod-model', a simple transport configuration where neutrons are constrained to move along a line, the only permissible directions being those in the increasing or decreasing spatial coordinate \cite{wing1962introduction}. Within this framework, we establish reference solutions for the asymptotic average number of neutrons stemming from an ancestor particle, as well as for the second moment of this counting process, for sub-critical, critical and super-critical configurations. These findings can be used as an ideal verification test-bed for Monte Carlo code developers interested in IFP-based algorithms.

This paper is organized as follows. In Sec.~\ref{sec:setting}, we introduce the general neutron transport model that we will work with and present a stochastic representation of the model in the form of a branching process. In Sec.~\ref{sec:statements}, we give an informal statement of the main results. In Sec.~\ref{sec:example} we consider the specific example of the rod model to illustrate the agreement between our theoretical results and Monte Carlo simulations in some simple cases. Conclusions will finally be drawn in Sec.~\ref{sec:conclusions}. The precise statements of the main results described in Sec.~\ref{sec:statements} are provided in the Appendix (Sec.~\ref{sec:appendix}) with the proofs being given in the supplementary material of Ref.~\citenum{NTE-mom-sup}.

\section{The stochastic interpretation of the neutron transport equation}
\label{sec:setting}

Consider a collection of neutrons evolving in the position (${\bf r}$), direction (${\mathbf \Omega}$) and energy ($E$) phase space $\mathcal S := D \times \mathbb S_2 \times (E_\mathtt{min}, E_\mathtt{max})$, where the spatial domain $D \subset \mathbb{R}^3$ is open and bounded, the direction domain is the unit sphere $\mathbb S_2 \subset \mathbb{R}^3$, and the energy domain is $(E_\mathtt{min}, E_\mathtt{max})$, where $0 < E_\mathtt{min} \le E_\mathtt{max} < \infty$.

It is customarily assumed that neutron trajectories behave stochastically, their randomness being due to the interactions with the surrounding nuclei; furthermore, thanks to the very low density of neutrons in matter, their transport process is inherently linear, in that the probability of neutron-neutron interactions is negligible for all practical purposes. Between collisions, particles move along straight lines whose length obeys a non-homogeneous exponential probability distribution with parameter $\Sigma_\mathtt{t}({\bf r}, E)$, with units given by inverse length. The total macroscopic cross section $\Sigma_\mathtt{t}({\bf r}, E)$ defines the probability per unit length that the neutron has a collision in the following infinitesimal displacement about its current position ${\bf r}$, in the direction ${\mathbf \Omega}$ and with energy $E$. At the scale of neutron paths, materials are ideally isotropic, so that $\Sigma_\mathtt{t}$ does not depend on ${\mathbf \Omega}$. Displacements are formally associated to the backward streaming operator
\begin{equation}
\Tdag[g]({\bf r}, {\mathbf \Omega}, E) = - {\mathbf \Omega} \cdot \nabla g({\bf r}, {\mathbf \Omega}, E) + \Sigma_\mathtt{t}({\bf r}, E)g({\bf r}, {\mathbf \Omega}, E), \quad ({\bf r}, {\mathbf \Omega}, E) \in \mathcal S,
\label{eq:transport}
\end{equation}
with $g \in L^\infty_+(\mathcal S)$ the collection of non-negative, uniformly bounded, measurable functions on $\mathcal S$.

We assume that the domain $D$ has leakage boundary conditions, which means that neutrons leaving the domain are lost. At the end of each flight, the neutron undergoes a collision event, provided it is still within the viable domain. We have $\Sigma_\mathtt{t}= \Sigma_\mathtt{c} +  \Sigma_\mathtt{s} +  \Sigma_\mathtt{f}$, where $\Sigma_\mathtt{c}$ (resp.  $\Sigma_\mathtt{s}$,  $\Sigma_\mathtt{f}$) denotes the capture (resp. scattering, fission) cross section. The ratios $\Sigma_\mathtt{r} / \Sigma_\mathtt{t}$, with $\mathtt{r}=\mathtt{c}$ (resp. $\mathtt{s}$, $\mathtt{f}$), yield the probability that the collision event is capture (resp. scattering or fission). If the neutron is captured, its history is terminated; if it undergoes scattering, its direction and energy coordinates are randomly distributed according to a (normalized) probability density $f_\mathtt{s}({\bf r}, {\mathbf \Omega} \to {\mathbf \Omega}', E \to E')$; if it undergoes fission, it disappears and is replaced by a random number of new neutrons, whose average number is $\nu_\mathtt{f}({\bf r}, E)$ and whose direction and energy coordinates are randomly distributed according to a (normalized) probability density $\chi_\mathtt{f}({\bf r}, {\mathbf \Omega} \to {\mathbf \Omega}', E \to E')$. Scattering and fission are formally associated to the backward operators
\begin{align}
\Sdag [g]({\bf r}, {\mathbf \Omega}, E) &= \Sigma_\mathtt{s}({\bf r}, E) \int_{E_\mathtt{min}}^{E_\mathtt{max}}\int_{\mathbb S_2} g({\bf r}, {\mathbf \Omega}', E') f_\mathtt{s}({\bf r}, {\mathbf \Omega} \to {\mathbf \Omega}', E \to E')\d {\mathbf \Omega}' \d E', \label{scattering}\\
\Fdag [g]({\bf r}, {\mathbf \Omega}, E) &= \nu_\mathtt{f}({\bf r}, E) \Sigma_\mathtt{f}({\bf r}, E) \int_{E_\mathtt{min}}^{E_\mathtt{max}}\int_{\mathbb S_2} g({\bf r}, {\mathbf \Omega}', E') \chi_\mathtt{f}({\bf r}, {\mathbf \Omega} \to {\mathbf \Omega}', E \to E')\d {\mathbf \Omega}' \d E' \label{fission}.
\end{align}
For now, in order to keep the notation simple, we only focus on prompt neutrons however, we refer the reader to Remark \ref{rem:delayed} for an explanation of how this is easily extended to the case where delayed neutrons are also accounted for.

The backward operators $\Tdag$, $\Sdag$ and $\Fdag$ are related to each other via the $k$-eigenvalue formulation of the adjoint neutron transport equation (NTE) \cite{LuxKoblinger}, which reads
\begin{equation}
(\Tdag - \Sdag)[\varphi^\dagger]= \frac{1}{k} \Fdag[\varphi^\dagger] \quad \Leftrightarrow \quad (\Tdag - \Sdag)^{-1}\Fdag[\varphi^\dagger] = k\varphi^\dagger. 
\label{eq:eval-prob}
\end{equation}
Taking the adjoint of all the operators in Eq.~\eqref{eq:eval-prob}, we have the customary forward formulation
\begin{equation}
(T - S)[\varphi]= \frac{1}{k} F[\varphi] \quad \Leftrightarrow \quad (T - S)^{-1}F[\varphi] = k\varphi,
\label{eq:adjoint}
\end{equation}
where $T$, $S$ and $F$ are the formal adjoints of the operators $\Tdag$, $\Sdag$ and $\Fdag$, respectively. The fundamental eigenmodes $\varphi^\dagger_0({\bf r}, {\mathbf \Omega}, E)$ and $\varphi_0({\bf r}, {\mathbf \Omega}, E)$ of Eqs.~\eqref{eq:eval-prob} and~\eqref{eq:adjoint}, respectively, can be given a physical meaning: $\varphi^\dagger_0({\bf r}, {\mathbf \Omega}, E)$ is the importance of a neutron injected into the system with phase-space coordinates $({\bf r}, {\mathbf \Omega}, E)$, and $Q_\mathtt{f} = F[\varphi_0]$ is the stationary {\it post-fission} neutron distribution (with $\varphi_0$ being the {\it pre-fission} steady state). We will expand on this further in due course. Since $\phidag_0$ and $Q_\mathtt{f}$ are unique up to multiplicative constants, we specify their normalisation as
\begin{equation}
  \langle Q_\mathtt{f}, \mathbf{1} \rangle = 1 \quad \text{ and }\quad \langle Q_\mathtt{f}, \phidag_0 \rangle = 1,
  \label{eq:norm}
\end{equation}
where $\mathbf{1}$ is the function that takes the value $1$ everywhere and the angle brackets denote the scalar product. The physical picture related to Eqs.~\eqref{eq:eval-prob} and~\eqref{eq:adjoint} is the following: given a population of pre-fission particles distributed according to $\varphi_0$, $F$ initiates the next generation of fission particles and $(T - S)^{-1}$ transports them to the next set of absorption (capture plus fission) sites, resulting in another population of particles distributed according to $\varphi_0$, albeit multiplied by a factor of $k_0$. In this respect, it is natural to introduce the concept of fission generation, and rewrite Eq.~\eqref{eq:adjoint} as
\begin{equation}
  F(T - S)^{-1}Q_\mathtt{f} = k_0 Q_\mathtt{f},
  \label{eq:forward_Q}
\end{equation}
where the fundamental eigenvalue $k_0$ physically represents the ratio between the number of neutrons in two successive fission generations. If $k_0 > 1$ the system is supercritical, corresponding to an exponential growth of the number of particles in system with respect to generations; if $k_0 < 1$ the system is subcritical, corresponding to an exponential decay in the number of particles; and if $k_0 = 1$ there is an equilibrium between neutron loss (absorption and leakage) and gain (fission).

The probabilistic interpretation of Eqs.~\eqref{eq:eval-prob} and~\eqref{eq:adjoint} is now made explicit. Let $N_n$ denote the number of neutrons that stem from the $n$-th fission event in their genealogical line of descent, with $n \ge 1$; furthermore, let $\{({\bf r}_i^{(n)}, {\mathbf \Omega}_i^{(n)}, E_i^{(n)}), \, i = 1, \dots, N_n\}$ denote their phase space configurations. Our goal is to characterize the statistical behaviour of the collection of neutrons in the $n$-th fission generation, which, in view of the previous remarks, is clearly a discrete-time branching process. Define
\begin{equation}
  \mathcal{X}_n(A) := \sum_{i = 1}^{N_n} \delta_{({\bf r}_i^{(n)}, {\mathbf \Omega}_i^{(n)}, E_i^{(n)})}(A), \qquad A \in \mathcal{B}(\mathcal S), n \ge 1,
  \label{eq:X}
\end{equation}
where $\mathcal{B}(\mathcal S)$ is the set of Borel subsets of $\mathcal S$. Then $\mathcal X_n(A)$ denotes the number of post-fission neutrons in $A \subset\mathcal S$. Moreover, $\mathcal X = (\mathcal{X}_n)_{n \ge 1}$ is a branching process with expectation semigroup
\begin{equation}
   \Psi_n[f]({\bf r}, {\mathbf \Omega}, E) := \mathbb{E}_{\delta_{({\bf r}, {\mathbf \Omega}, E)}}\left[\sum_{i = 1}^{N_n}f({\bf r}_i^{(n)}, {\mathbf \Omega}_i^{(n)}, E_i^{(n)})\right], 
\end{equation}
where $\mathbb E_{\delta_{({\bf r}, {\mathbf \Omega}, E)}}$ is the expectation operator associated with the law $\mathbb P_{\delta_{({\bf r}, {\mathbf \Omega}, E)}}$ of the branching process when initiated from a single particle at $({\bf r}, {\mathbf \Omega}, E) \in \mathcal S$ and $f\in L_+^\infty(\mathcal S)$, the set of non-negative, bounded measurable functions on $\mathcal S$. The physical interpretation of $\Psi_n[f]({\bf r}, {\bf \Omega}, E)$ is the expected behaviour of $\mathcal X_n$ when initiated from a single neutron with configuration $({\bf r}, {\mathbf \Omega}, E) \in \mathcal S$. For example, taking $f = 1$, then $\Psi_n[f]({\bf r}, {\bf \Omega}, E)$ is equal to $\mathbb E_{\delta_{({\bf r}, {\bf \Omega}, E)}}[N_n]$, the expected number of $n$-th generation post-fission neutrons, which has direct bearing on the IFP algorithm. 

\section{Statistics of fission chains: main results}
\label{sec:statements}

It is well known, see e.g. Ref.~\citenum{SNTE3}, that $\Psi_n[f]$ is related to $k_0$, $\varphi_0^\dagger$ and $Q_\mathtt{f}$ by
\begin{equation}
    \Psi_n[\varphi_0^\dagger] = k_0^n \varphi_0^\dagger, \quad \text{ and } \quad 
  \langle Q_\mathtt{f}, \Psi_n[f]\rangle = k_0^n \langle Q_\mathtt{f}, f\rangle,
\end{equation}
where brackets denote again the scalar product. Furthermore, it is also known that, see also Ref. \citenum{SNTE3}, 
\begin{equation}
 \Psi_n[f]({\bf r}, {\mathbf \Omega}, E) \sim k_0^{n} \langle Q_\mathtt{f}, f\rangle \varphi_0^\dagger({\bf r}, {\mathbf \Omega}, E),
\label{eq:PF}
\end{equation}
for large $n$. In particular, taking $f = 1$ in the above asymptotic and using Eq. \eqref{eq:norm}, we obtain
\begin{equation}
  \mathbb E_{\delta_{({\bf r, \bf \Omega, E})}}[N_n] \sim k_0^n \varphi_0^\dagger({\bf r}, {\bf \Omega}, E),
  \label{eq:PFsimple}
\end{equation}
which shows that the expected number of particles in the system for large $n$ is proportional to $\varphi_0^\dagger({\bf r}, {\bf \Omega}, E)$, demonstrating that $\varphi_0^\dagger$ indeed physically represents the importance of an initial particle at $({\bf r}, {\bf \Omega}, E)$ to the growth of the neutron population in the system.  

In addition to the asymptotic behaviour of the first moment given in Eq.~\eqref{eq:PF}, in the last decade several results pertaining to the asymptotic behaviour of moments of functionals of $(\mathcal{X}_n)_{n \ge 1}$ have emerged, see for example Refs.~\citenum{zoia2011residence, zoia2012counting, zoia2012discrete} and Refs.~\citenum{moments, Schertzer} for general branching processes. In the present article, we unify and extend the aforementioned results in the spirit of Ref.~\citenum{moments}, which deals with the evolution of a neutron population described in terms of a branching process in a continuous-time setting. More precisely, for $n, \ell \ge 1, f \in L^\infty_+(\mathcal S)$ and $({\bf r}, {\mathbf \Omega}, E) \in \mathcal S$, define
\begin{equation}
     \Psi_n^{(\ell)}[f]({\bf r}, {\mathbf \Omega}, E) := \mathbb{E}_{\delta_{({\bf r}, {\mathbf \Omega}, E)}}\left[\left(\sum_{i = 1}^{N_n}f({\bf r}_i^{(n)}, {\mathbf \Omega}_i^{(n)}, E_i^{(n)})\right)^\ell\right]. 
\end{equation}
We will show that, regardless of the value of $k_0$, for each $\ell \ge 1$ there exist $g_\ell(n)$ and $h_\ell$ such that 
\begin{equation}
 \Psi_n^{(\ell)}[f]({\bf r}, {\mathbf \Omega}, E) \sim  \frac{h_\ell}{g_\ell(n)} \varphi_0^\dagger({\bf r}, {\mathbf \Omega}, E)
\label{eq:thm*}
\end{equation}
for large $n$. Note that, thanks to Eq.~\eqref{eq:PF}, in the case $\ell = 1$ we have $g_1(n) = k_0^{-n}$ and $h_1 = \langle Q_\mathtt{f}, f\rangle$.

Our second contribution will be to show that, in the critical case when $k_0=1$, one can obtain a precise asymptotic for the survival probability, $\mathbb{P}_{\delta_{({\bf r}, {\mathbf \Omega}, E)}}(N_n > 0)$ for a population descending from a single particle starting at $({\bf r}, {\mathbf \Omega}, E)$. This result, combined with the moment asymptotics in the critical case, implies that the limiting distribution of $\frac1n\sum_{i = 1}^{N_n}f({\bf r}_i^{(n)}, {\mathbf \Omega}_i^{(n)}, E_i^{(n)})$, conditional on $N_n > 0$, is exponential with a rate that can be explicitly determined by the model parameters. The latter two results are classical in the branching processes literature: the survival probability dates back to Kolmogorov, see Ref.~\citenum{Kol}, and the second result is due to Yaglom, see Ref.~\citenum{Yaglom}. We also refer the reader to Refs.~\citenum{YaglomNTE, Ellen} for continuous-time results, and to Refs.~\citenum{geiger, mullikin} for discrete-time results. The proofs of our derivations are provided in Sec.~\ref{sec:appendix} and in the supplementary material given in Ref.~\citenum{NTE-mom-sup}. In the case where one assumes reflective boundary conditions instead of leakage, the moment asymptotic results (for all values of $k_0$) would still hold as long as Eq.~\eqref{eq:PF} is true: this is due to the fact that the proofs are inductive, so only require the result to be true for the first moment. 

\subsection{The critical case}

For the critical case, with $k_0=1$, it is possible to show (see Sec.~\ref{sec:appendix}) that
\begin{equation}\label{eq:crit-mom}
\Psi_n^{(\ell)}[f]({\bf r}, {\mathbf \Omega}, E)
\sim n^{\ell-1}\frac{\ell!}{2^{\ell-1}} \,\langle Q_\mathtt{f}, f\rangle^\ell
  \langle\varphi_0, \Sigma_{\mathtt f}\mathcal{V}[\varphi_0^\dagger]\rangle^{\ell-1}   \varphi_0^\dagger({\bf r}, {\mathbf \Omega}, E)
\end{equation}
for sufficiently large $n$, where for $g \in L^\infty_+(\mathcal S)$ we have defined
\begin{equation}
\label{eq:V}
\mathcal{V}[g]({\bf r}, {\mathbf \Omega}, E) 
= \mathcal{E}_{({\bf r}, {\mathbf \Omega}, E)}\left[\left(\sum_{i = 1}^M g({\bf r}, {\mathbf \Omega}_i, E_i)\right)^2 \right] - \mathcal{E}_{({\bf r}, {\mathbf \Omega}, E)}\left[ \sum_{i = 1}^M g^2({\bf r}, {\mathbf \Omega}_i, E_i)\right],
\end{equation}
where $M$ denotes the random number of neutrons produced from a fission event and $\mathcal{E}_{({\bf r}, {\mathbf \Omega}, E)}$ is the average over the fission offspring (number and configurations) produced from a fission event at $({\bf r}, {\mathbf \Omega}, E)$.

To illustrate Eq. \eqref{eq:crit-mom} further, we consider some simples cases. For example, when $\ell = 1$, we see that we recover Eq. \eqref{eq:PF} and hence, in particular, taking $f = 1$, we recover Eq. \eqref{eq:PFsimple}. Now let us consider the case when $f = 1$ and $\ell = 2$. In this case, thanks to Eq. \eqref{eq:norm}, Eq. \eqref{eq:crit-mom} becomes
\begin{equation}\label{eq:crit_mom2}
\mathbb E_{\delta_{({\bf r}, {\mathbf \Omega}, E)}}[N_n^2]
\sim  
  n\langle\varphi_0, \Sigma_{\mathtt f}\mathcal{V}[\varphi_0^\dagger]\rangle  \varphi_0^\dagger({\bf r}, {\mathbf \Omega}, E),
\end{equation}
for large $n$, which means that the second moment diverges linearly with respect to the fission generations. 

As promised, for the critical case, we also give a precise asymptotic for the survival probability. Theorem \ref{thm:survival} of Sec.~\ref{sec:appendix} shows that for any $({\bf r}, {\mathbf \Omega}, E) \in \mathcal S$ we have
\begin{equation}
\mathbb{P}_{\delta_{({\bf r}, {\mathbf \Omega}, E)}}(N_n > 0) \sim \frac{2}{n} \frac{\varphi_0^\dagger({\bf r}, {\mathbf \Omega}, E)}{\langle \varphi_0, \Sigma_{\mathtt f} \mathcal{V}[\varphi_0^\dagger]\rangle}
\label{eq:survival}
\end{equation}
as $n \to \infty$, where $\mathcal V$ was defined in Eq.~\eqref{eq:V}. Here we see another perspective of the aforementioned interpretation of $\varphi_0^\dagger$ in the critical case, in that the survival probability is proportional to $\varphi_0^\dagger$.

Combining Eqs. \eqref{eq:crit-mom} and \eqref{eq:survival}, it is straightforward to show that for any $({\bf r}, {\mathbf \Omega}, E) \in \mathcal S$ and any $f \in L^\infty_+(\mathcal S)$, conditional on survival, the distribution of the process normalised by the current generation $n$ is asymptotically exponential. That is,  
\begin{equation}
\left(\frac1n \sum_{i = 1}^{N_n}f({\bf r}_i^{(n)}, {\mathbf \Omega}_i^{(n)}, E_i^{(n)}) \Bigg| N_n > 0\right) \sim Y
\label{eq:Yaglom}
\end{equation}
for large $n$, where $Y$ is an exponential random variable with rate $2/\langle Q_\mathtt{f}, f\rangle \langle \varphi_0, \Sigma_{\mathtt f}\mathcal V[\varphi_0^\dagger]\rangle$. We refer the reader to Ref. \cite{NTE-mom-sup} for the details of the proof.

\subsection{The supercritical case}

For the supercritical case, with $k_0>1$, the findings of Sec.~\ref{sec:appendix} show that
\begin{equation}\label{eq:sup-mom}
\Psi_n^{(\ell)}[f]({\bf r}, {\mathbf \Omega}, E) \sim k_0^{n\ell} \,\langle Q_\mathtt{f}, f \rangle^\ell L_\ell({\bf r}, {\mathbf \Omega}, E)  \, \varphi_0^\dagger({\bf r}, {\mathbf \Omega}, E)
\end{equation}
for large $n$. Here $L_\ell({\bf r}, {\mathbf \Omega}, E)$ is defined recursively with $L_1 = 1$ and for $\ell \ge 2$,
\begin{equation}
L_\ell({\bf r}, {\mathbf \Omega}, E) 
= \frac{{\ell!}}{\varphi_0^\dagger({\bf r}, {\bf \Omega}, E)}\sum_{m =1}^\infty \frac{1}{k_0^{m\ell}}
\Psi_m^-\Bigg[\mathcal{E}_\cdot\Bigg[\sum_{[\ell_1, \dots, \ell_M]_\ell^{2+}}\prod_{\substack{j = 1 \\ \ell_j > 0}}^M L_{\ell_j}(\cdot, {\mathbf \Omega}_j, E_j)\varphi_0^\dagger(\cdot, {\mathbf \Omega}_j, E_j)\Bigg]\Bigg]({\bf r}, {\mathbf \Omega}, E),
\label{eq:supL}
\end{equation}
where $[\ell_1, \dots, \ell_M]_\ell^{2+}$ denotes the set of non-negative tuples $(\ell_1, \dots, \ell_M)$, such that $\sum_{j = 1}^M \ell_j = \ell$ and at least two of the $\ell_j$ are strictly positive. We have used the notation $\Psi_n^-$ for the semigroup associated to the collection of particles stopped {\em just before} the $n$-th fission event in their genealogical lines of descent.

Again, in the case where $\ell = 1$, we recover Eq.~\eqref{eq:PFsimple}. Now considering the case where $\ell = 2$ and $f = 1$, we have
\begin{equation}\label{eq:supmom2}
    \mathbb E_{\delta_{({\bf r}, {\bf \Omega}, E)}}[N_n^2] \sim k_0^{2n}L_2({\bf r}, {\bf \Omega}, E)\varphi_0^\dagger(\bf r, \bf \Omega, E),
\end{equation}
where we have used the fact that $\langle Q_{\mathtt f}, {\bf 1}\rangle = 1$ from Eq.~\eqref{eq:norm}. Since $L_1 = 1$, it follows that $L_2$ is given by
\begin{equation}\label{eq:L2sup_mom2}
    L_2({\bf r}, {\bf \Omega}, E) = \frac{2}{\varphi_0^\dagger({\bf r}, {\bf \Omega}, E)} \sum_{m = 1}^\infty \frac{1}{k_0^{2m}}\Psi_m^-\Bigg[\mathcal{E}_\cdot\Bigg[\sum_{[\ell_1, \dots, \ell_M]_2^{2+}}\prod_{\substack{j = 1 \\ \ell_j > 0}}^M \varphi_0^\dagger(\cdot, {\mathbf \Omega}_j, E_j)\Bigg]\Bigg]({\bf r}, {\mathbf \Omega}, E).
\end{equation}

\subsection{The subcritical case}

For the subcritical case, with $k_0<1$, the findings of Sec.~\ref{sec:appendix} show that
\begin{equation}\label{eq:sub-mom}
\Psi_n^{(\ell)}[f]({\bf r}, {\mathbf \Omega}, E) \sim k_0^{n}\, L_\ell \, \varphi_0^\dagger({\bf r}, {\mathbf \Omega}, E)
\end{equation}
for large $n$, where again $L_\ell$ is defined recursively, with $L_1 = \langle Q_\mathtt{f}, f\rangle$ and
\begin{equation}
  L_\ell = \langle Q_\mathtt{f},  f^\ell \rangle + \sum_{m = 0}^\infty \frac{1}{k_0^m}\left\langle \varphi_0 , \Sigma_{\mathtt f}(\cdot)\mathcal{E}_\cdot\Bigg[\sum_{[\ell_1, \dots, \ell_M]_\ell^{2+}}{\ell \choose \ell_1 \dots \ell_M}\prod_{{\substack{j = 1 \\ \ell_j > 0}}}^M\Psi_m^{(\ell_j)}[f](\cdot, {\mathbf \Omega}_j, E_j)\Bigg]\right\rangle,  
\end{equation}
for $\ell \ge 2$. The set $[\ell_1, \dots, \ell_M]_\ell^{2+}$ is defined as in the previous section.

As with the other two cases, setting $\ell = 1$ we recover Eq.~\eqref{eq:PF}. Setting $f = 1$ and $\ell = 2$ and using Eq.~\eqref{eq:norm}, we have
\begin{equation}\label{eq:submom2}
    \mathbb E_{\delta_{({\bf r}, {\bf \Omega}, E)}}[N_n^2] \sim k_0^n L_2 \varphi_0^\dagger({\bf r}, {\bf \Omega}, E),
\end{equation}
where
\begin{equation}
  L_2 = 1 + 2\sum_{m = 0}^\infty \frac{1}{k_0^m}\left\langle \varphi_0 , \Sigma_{\mathtt f}(\cdot)\mathcal{E}_\cdot\Bigg[\sum_{[\ell_1, \dots, \ell_M]_2^{2+}}\prod_{{\substack{j = 1 \\ \ell_j > 0}}}^M \mathbb E_{\delta_{(\cdot, {\mathbf \Omega}_j, E_j)}}[N_m]\Bigg]\right\rangle. 
\label{eq:L2sub_mom2}
\end{equation}

\begin{rem}\label{rem:delayed}
    In the case where we consider delayed neutron production, as well as prompt neutrons, the fission operator $\Fdag$ would be replaced by 
    \begin{align*}
    \Fdag [g]({\bf r}, {\mathbf \Omega}, E) &= \nu_\mathtt{f,p}({\bf r}, E) \Sigma_\mathtt{f}({\bf r}, E) \int_{E_\mathtt{min}}^{E_\mathtt{max}}\int_{\mathbb S_2} g({\bf r}, {\mathbf \Omega}', E') \chi_\mathtt{f,p}({\bf r}, {\mathbf \Omega} \to {\mathbf \Omega}', E \to E')\d {\mathbf \Omega}' \d E'\\ 
    &\quad +\sum_{i=1}^D \nu_\mathtt{f,d,i}({\bf r}, E) \Sigma_\mathtt{f}({\bf r}, E) \int_{E_\mathtt{min}}^{E_\mathtt{max}}\int_{\mathbb S_2} g({\bf r}, {\mathbf \Omega}', E') \chi_\mathtt{f,d,i}({\bf r}, {\mathbf \Omega} \to {\mathbf \Omega}', E \to E')\d {\mathbf \Omega}' \d E',
    \end{align*}
    where $\nu_\mathtt{f,p}({\bf r}, E)$ denotes the average number of prompt neutrons produced at a fission event whose directions and energies are randomly distributed according to a (normalised) probability density $\chi_\mathtt{f,p}({\bf r}, {\mathbf \Omega} \to {\mathbf \Omega}', E \to E')$, and similarly $\nu_\mathtt{f,d,i}({\bf r}, E)$ denotes the average number of delayed neutrons coming from the $i$-th precursor, whose directions and energies are distributed according to $\chi_\mathtt{f,d,i}({\bf r}, {\mathbf \Omega} \to {\mathbf \Omega}', E \to E')$, for $i = 1, \dots, D$. Since, in generational time, both prompt and delayed neutrons are produced on the same time scales, our results remain unchanged. More precisely, we could let $M_i$ denote the random number of delayed neutrons produced from the $i$-th precursor and $M_0$ denote the number of prompt neutrons produced, so that our hold in the setting of delayed neutrons by taking $M = M_0 + \dots + M_D$.
\end{rem}

\section{Rod model}
\label{sec:example}

In this section we illustrate the main statements of Sec.~\ref{sec:statements} by considering the one-dimensional rod model, which is a highly simplified neutron transport configuration where analytical results can be easily established \cite{Zoia2012branching}. As such, the rod model provides an ideal benchmark framework for the verification of Monte Carlo simulations; contrary to the infinite-medium two-group setup, the rod model in particular offers the possibility of effectively probing spatial and angular effects.

The rod model, which was introduced by the pioneering work by G.~M.~Wing \cite{wing1962introduction}, assumes that neutron displacements are restricted to a straight line, with the only permissible directions being $\{-1, +1\}$. For the benchmark considered here, we take the viable spatial domain to be the bounded segment $D = (-R, R)$ for some $R > 0$, so that the spatial position of the particle can be described using the scalar coordinate $x \in  (-R, R)$. For the sake of simplicity, we will further assume that all neutrons have unit energy. Thus, in this case, the phase space is reduced to $\mathcal S = D \times \{-1, +1\}$. Moreover, we impose leakage boundary conditions at $x=-R$ and $x=R$.

As for the nuclear data, we take spatially-homogeneous cross sections
\begin{align}
  &\qquad\Sigma_\mathtt{s}(x) \equiv \Sigma_\mathtt{s} > 0, \quad  
  \Sigma_\mathtt{f}(x) \equiv \Sigma_\mathtt{f} > 0, \quad   
  \Sigma_\mathtt{c}(x) \equiv \Sigma_\mathtt{c} > 0.
  \label{eq:cross-sections}
\end{align}
The scattering and fission distributions are assumed to be isotropic:
\begin{align}
  &f_\mathtt{s}(x, \Omega\to \Omega') = \chi_\mathtt{f}(x, \Omega\to \Omega') = \frac{1}{2} \left(\delta(\Omega - \Omega') + \delta(\Omega + \Omega')\right).
  \label{eq:kernels}
\end{align}
Finally, we assume that exactly two particles are emitted at fission events, which imposes the average fission multiplicity $\nu_\mathtt{f} = 2$.

For this choice of the physical parameters, the adjoint operators $\Sdag$ and $\Fdag$ are given by
\begin{align}
\Sdag [g](x, \Omega) &= \frac{\Sigma_\mathtt{s}}{2}\big(g(x, \Omega) + g(x, -\Omega)\big)\\
\Fdag [g](x, \Omega) &= \frac{\nu_\mathtt{f} \Sigma_\mathtt{f}}{2}\big(g(x, \Omega) + g(x, -\Omega)\big) = \Sigma_\mathtt{f}\big(g(x, \Omega) + g(x, -\Omega)\big) .
\end{align}
Correspondingly, the eigenvalue problem in Eq.~\eqref{eq:eval-prob} yields the set of two coupled ordinary differential equations (ODE) 
\begin{align}
-\Omega &\frac{d}{d x}\varphi^\dagger(x, \Omega) + \Sigma_\mathtt{t}\varphi^\dagger(x, \Omega) - \frac{\Sigma_\mathtt{s}}{2}\big(\varphi^\dagger(x, \Omega) + \varphi^\dagger(x, -\Omega)\big) = \frac{\Sigma_\mathtt{f}}{k}(\varphi^\dagger(x, \Omega) + \varphi^\dagger(x, -\Omega)\big) ,
\label{eq:eval-prob-eg}
\end{align}
for $\Omega = \pm 1$, with the boundary conditions
\begin{align}
&\varphi^\dagger(R, \Omega= +1) = \varphi^\dagger(-R, \Omega=-1) = 0.
\label{eq:bc}
\end{align}
Standard ODE techniques (see e.g.~Refs.~\citenum{MCNTE, Zoia2012branching}) show that $k_0$ is the largest $k$ that satisfies the dispersion relation
\begin{equation}
  2 \cos(2 R \alpha_k) = \left(\frac{\alpha_k }{\Sigma_\mathtt{t}}- \frac{\Sigma_\mathtt{t}}{\alpha_k} \right) \sin(2 R \alpha_k), 
\end{equation}
where
\begin{equation}
  \alpha_k = \sqrt{\Sigma_\mathtt{t}(\nu_{\mathtt f} \Sigma_{\mathtt f} / k + \Sigma_{\mathtt s} - \Sigma_\mathtt{t})}.
\end{equation}
Based on the dispersion law for $k$, it is possible to choose a combination of nuclear data and system size such that $k_0$ is equal to some target value: for a given $k_0$, we define
\begin{equation}
  {\bar c}_0 = \frac{\frac{\nu_{\mathtt f} \Sigma_{\mathtt f}}{k_0} +  \Sigma_{\mathtt s}}{ \Sigma_{\mathtt t}},
\end{equation}
with $\Sigma_{\mathtt t} = \Sigma_{\mathtt c} + \Sigma_{\mathtt s} + \Sigma_{\mathtt f}$, and the corresponding value of the system size must satisfy
\begin{equation}
  R_0 = \frac{\arctan\left(\frac{1}{\sqrt{{\bar c}_0 -1}} \right)}{\Sigma_{\mathtt t}\sqrt{{\bar c}_0 -1}}.
  \label{eq:rod_L0}
\end{equation}
In particular, when $k_0=1$, the choice of $R_0$ corresponds to the critical half-size of the rod. Since $k_0 < k_\infty := \nu_{\mathtt f} \Sigma_{\mathtt f} / (\Sigma_{\mathtt c} + \Sigma_{\mathtt f})$, $k_\infty$ being the infinite multiplication factor, a necessary condition for the existence of a $R_0$ ensuring $k_0 \ge 1$ is that the nuclear data are chosen so that $ k_\infty > 1$.

The adjoint and forward dominant eigenmodes are readily obtained, recalling that we impose the normalisation $\langle Q_\mathtt{f}, \mathbf{1}\rangle = 1$ and $\langle Q_\mathtt{f}, \varphi_0^\dagger\rangle = 1$, which fixes the multiplicative constants. For the adjoint dominant eigenfunction we have
\begin{equation}
  \varphi_0^\dagger(x, \Omega=\pm 1) =\frac{8 \sin(\alpha_{k_0} R_0)}{4 R_0 \alpha_{k_0} + 2 \sin(2 \alpha_{k_0} R_0)}\left[\cos(\alpha_{k_0} x) \mp \frac{\sin(\alpha_{k_0} x)}{\tan(\alpha_{k_0} R_0)} \right],
  \label{eq:rod_adj_model}
\end{equation}
whereas for the forward dominant eigenfunction we have
\begin{equation}
  \varphi_0(x, \Omega=\pm 1) = \frac{\alpha_{k_0}}{4 \nu_{\mathtt f} \Sigma_{\mathtt f} \sin(\alpha_{k_0} R_0)} \left[\cos(\alpha_{k_0} x) \pm \frac{\sin(\alpha_{k_0} x)}{\tan(\alpha_{k_0} R_0)} \right],
  \label{eq:rod_for_model}
\end{equation}
with
\begin{equation}
  Q_\mathtt{f}(x, \Omega) = \frac{\alpha_{k_0}}{4 \sin(\alpha_{k_0} R_0)} \cos(\alpha_{k_0} x) .
\end{equation}

\subsection{The statistical behavior of the fission chains}

Knowledge of the dominant eigenfunctions allows one to obtain explicit results for the asymptotic moments of the neutron population at large $n$. For the single-speed rod model, the semigroup of the branching process is given by 
\begin{equation}
  \Psi_n[f](x, \Omega) = \mathbb{E}_{\delta_{(x, \Omega)}}\left[\sum_{i = 1}^{N_n}f(x_i^{(n)}, \Omega_i^{(n)}) \right].
\label{psi-rod}
\end{equation}
In the following we will focus in particular on the case $f=1$, i.e., the counting process for the fission neutrons being in the rod at a given generation $n$.

\subsubsection{The critical case}

Let us consider the critical case, with $k_0 = 1$. Recalling that $N_n$ denotes the number of neutrons generated within the rod in the $n$-th fission generation, Eq.~\eqref{eq:PF} shows that
\begin{equation}
  \Psi_n[{\bf 1}](x, \Omega) = \mathbb{E}_{\delta_{(x, \Omega)}}[N_n] \sim \langle Q_\mathtt{f}, \mathbf{1} \rangle \varphi_0^\dagger(x, \Omega),
  \label{eq:1st-mom-limit}
\end{equation}
as $n \to \infty$. Since the normalization has been chosen so that $\langle Q_\mathtt{f}, \mathbf{1} \rangle = 1$, we therefore have 
\begin{equation}
\mathbb{E}_{\delta_{(x, \Omega)}}[N_n] \sim \varphi_0^\dagger(x, \Omega).
\end{equation}

\begin{figure}[t]
\centering
\includegraphics[scale = 0.7]{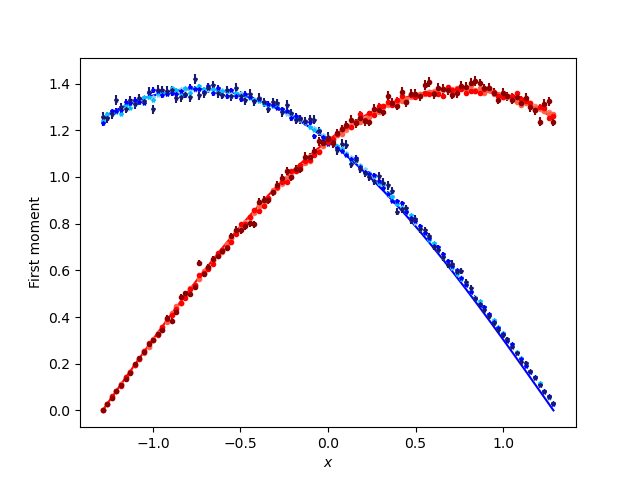}
\caption{Rod model benchmark for critical conditions, with $k_0=1$: analysis of the average number of fission neutrons. Comparison between the exact expression for $\varphi_0^\dagger(x, \Omega)$ in Eq.~\eqref{eq:1st-mom-limit} (solid lines) and Monte Carlo simulations for $\mathbb{E}_{\delta_{(x, \Omega)}}[N_n]$ (symbols). Error bars correspond to one sigma uncertainty. Shades of blue: case $\Omega=+1$, with increasing $n=10$, $n=50$, $n=100$, and $n=300$; shades of red: case $\Omega=-1$, with increasing $n=10$, $n=50$, $n=100$, and $n=300$.}
\label{fig:crit_N1}
\end{figure}

For illustration, in Fig.~\ref{fig:crit_N1} we display the comparison between the exact result in Eq.~\eqref{eq:1st-mom-limit} and Monte Carlo simulations. We have set the cross sections as
\begin{equation}
    \Sigma_{\mathtt s} = 0.3, \quad \Sigma_{\mathtt c} = 0.2, \quad \Sigma_{\mathtt f} = 0.7,
    \label{eq:rod_sigmas}
\end{equation}
with $2$ secondary neutrons per fission events, and we have imposed $D = (-R_0, R_0)$ where $R_0$ was given in Eq.~\eqref{eq:rod_L0} in order to ensure $k_0=1$. Monte Carlo simulations have been run using $10^6$ particles, with $n$ varying from $n=10$ to $n=300$. The Monte Carlo tally corresponds to the ensemble-averaged number of fission neutrons $\mathbb{E}_{\delta_{(x, \Omega)}}[N_n]$ being in the rod at generation $n$, for a single ancestor neutron starting with coordinates $(x,\Omega)$. The statistical agreement between Monte Carlo results and Eq.~\eqref{eq:1st-mom-limit} is very good. The asymptotic behavior is attained for relatively small values of $n$: at $n=10$, the average number of fission neutrons has already settled to the spatial shape given by Eq.~\eqref{eq:1st-mom-limit}.

\begin{figure}[t]
\centering
\includegraphics[scale = 0.7]{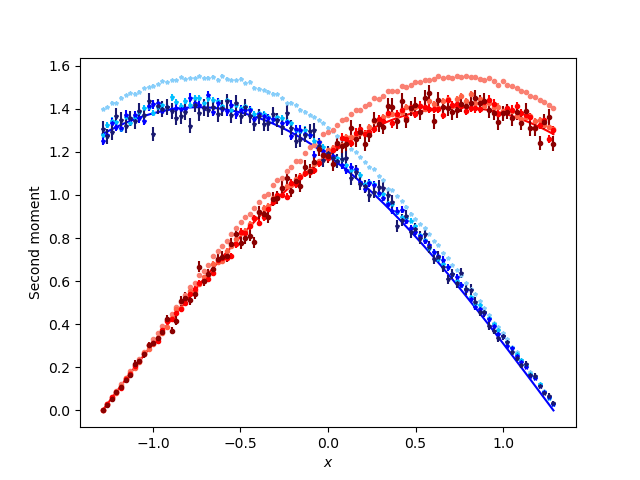}
\caption{Rod model benchmark for critical conditions, with $k_0=1$: analysis of the second moment of the number of fission neutrons. Comparison between the exact expression for $\Sigma_{\mathtt f} \langle \varphi_0, \mathcal V[\varphi_0^\dagger]\rangle \varphi_0^\dagger(x, \Omega)$ in Eq.~\eqref{eq:2nd-mom-limit} (solid lines) and Monte Carlo simulations for $\mathbb E_{\delta_{(x, \Omega)}}[N_n^2] / n$ (symbols). Error bars correspond to one sigma uncertainty. Shades of blue: case $\Omega=+1$, with increasing $n=10$, $n=50$, $n=100$, and $n=300$; shades of red: case $\Omega=-1$, with increasing $n=10$, $n=50$, $n=100$, and $n=300$.}
\label{fig:crit_N2}
\end{figure}

Now let us consider the second moment of the number of fission neutrons in the rod. Recall from Eq. \eqref{eq:crit_mom2} that
\begin{equation}
    \mathbb E_{\delta_{(x, \Omega)}}[N_n^2] \sim n\Sigma_{\mathtt f} \langle \varphi_0, \mathcal V[\varphi_0^\dagger]\rangle \varphi_0^\dagger(x, \Omega),
    \label{eq:2nd-mom-limit}
\end{equation}
as $n \to \infty$. Here we have used the fact that the system is spatially homogeneous, which implies that $\langle \Sigma_{\mathtt f} \varphi_0, \mathcal V[\varphi_0^\dagger]\rangle = \Sigma_{\mathtt f}\langle\varphi_0, \mathcal V[\varphi_0^\dagger]\rangle$. Moreover, combining Eq.~\eqref{eq:V} and the fact that fission is isotropic, we have
{\color{black}
\begin{align}
  \mathcal V[g](x, \Omega) 
  &= \mathcal{E}_{(x, \Omega)}[g^2(x, \Omega_1) + 2g(x, \Omega_1)g(x, \Omega_2) + g^2(x, \Omega_2)] - \mathcal{E}_{(x, \Omega)}[g^2(x, \Omega_1) + g^2(x, \Omega_2)] \notag\\
  &= 2\mathcal{E}_{(x, \Omega)}[ g(x, \Omega_1)g(x, \Omega_2)]\notag\\
  &= 2\mathcal{E}_{(x, \Omega)}[g(x, \Omega_1)]^2\notag\\
  &= \frac12\big[g(x, \Omega) + g(x, -\Omega)\big]^2.
  \label{eq:calc-V}
\end{align}
}
Based on this expression, we can easily derive
\begin{equation}
 \langle \varphi_0, \mathcal V[\varphi_0^\dagger]\rangle = \frac{16 \left[ 5 + \cos(2 \alpha_{k_0} R_0) \right] \sin^2(\alpha_{k_0} R_0)}{6 \Sigma_{\mathtt f} \left[2 \alpha_{k_0} R_0 + \sin(2 \alpha_{k_0} R_0) \right]^2}.
 \label{eq:phiVphiadj}
\end{equation}
For illustration, in Fig.~\ref{fig:crit_N2} we display the comparison between the exact result in Eq.~\eqref{eq:2nd-mom-limit}, normalized to $n$, and Monte Carlo simulations, also normalized to $n$. Cross sections and $R_0$ are the same as in the previous numerical example. Monte Carlo simulations have been run using $10^6$ particles, with $n$ varying from $n=10$ to $n=300$. The Monte Carlo tally corresponds to the ensemble-averaged normalized second moment $\mathbb{E}_{\delta_{(x, \Omega)}}[N^2_n]/n$ of the number of fission neutrons being in the rod at generation $n$, for a single ancestor neutron starting with coordinates $(x,\Omega)$. The statistical agreement between Monte Carlo results and Eq.~\eqref{eq:2nd-mom-limit} is again very good. The asymptotic behavior is attained later than the case of the average value: the second moment of fission neutrons is settled at $n=50$.

\begin{figure}[t]
\centering
\includegraphics[scale = 0.7]{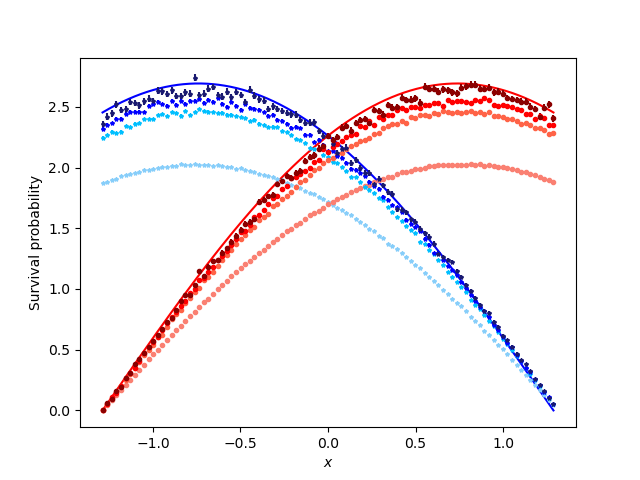}
\caption{Rod model benchmark for critical conditions, with $k_0=1$: analysis of the survival probability. Comparison between the exact expression for $2 \varphi_0^\dagger(x, \Omega) / (\Sigma_{\mathtt f}\langle \varphi_0,  \mathcal{V}[\varphi_0^\dagger]\rangle)$ in Eq.~\eqref{eq:rod_survival} (solid lines) and Monte Carlo simulations for $n \mathbb{P}_{\delta_{(x, \Omega)}}(N_n > 0)$ (symbols). Error bars correspond to one sigma uncertainty. Shades of blue: case $\Omega=+1$, with increasing $n=10$, $n=50$, $n=100$, and $n=300$; shades of red: case $\Omega=-1$, with increasing $n=10$, $n=50$, $n=100$, and $n=300$.}
\label{fig:rod_survival}
\end{figure}

Finally, the survival probability for the critical rod model can be explicitly computed using Eq.~\eqref{eq:survival}. Indeed, due to spatial homogeneity of the cross-sections, we have
\begin{equation}
    \mathbb{P}_{\delta_{(x, \Omega)}}(N_n > 0) \sim \frac{2}{n}\frac{\varphi_0^\dagger(x, \Omega)}{\Sigma_{\mathtt f}\langle \varphi_0,  \mathcal{V}[\varphi_0^\dagger]\rangle},
    \label{eq:rod_survival}
\end{equation}
which can be computed explicitly using Eqs. \eqref{eq:rod_adj_model} and \eqref{eq:phiVphiadj}. The comparison between Monte Carlo simulations and the exact formula in Eq.~\eqref{eq:rod_survival} is illustrated in Fig.~\ref{fig:rod_survival} as a function of the number of generations $n$. The convergence to the asymptotic shape requires a larger $n$ compared to the cases of the first and second moment of the number of fission neutrons: Monte Carlo simulations attain the shape of Eq.~\eqref{eq:rod_survival} at about $n=300$.

\subsubsection{The supercritical case}

\begin{figure}[h!]
\centering
\includegraphics[scale = 0.7]{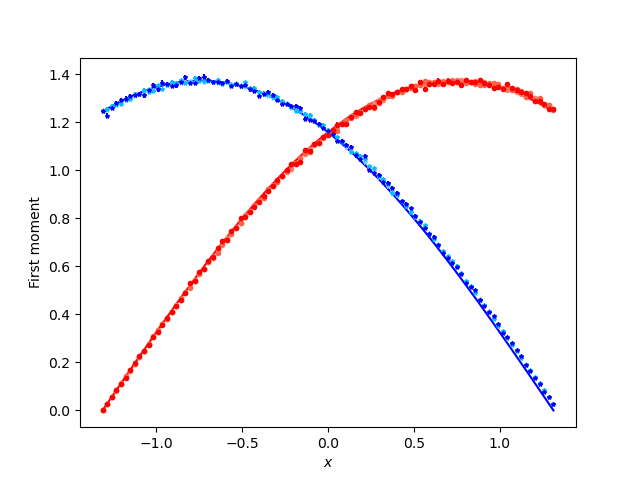}
\caption{Rod model benchmark for supercritical conditions, with $k_0=1.01$: analysis of the average number of fission neutrons. Comparison between the exact expression for $\varphi_0^\dagger(x, \Omega)$ in Eq.~\eqref{eq:PF2} (solid lines) and Monte Carlo simulations for $\mathbb{E}_{\delta_{(x, \Omega)}}[N_n] / k_0^{n}$ (symbols). Error bars correspond to one sigma uncertainty. Shades of blue: case $\Omega=+1$, with increasing $n=10$, $n=50$, and $n=100$; shades of red: case $\Omega=-1$, with increasing $n=10$, $n=50$, and $n=100$.}
\label{fig:super_N1}
\end{figure}

Next let us consider the supercritical case, with $k_0 > 1$. From Eqs.~\eqref{eq:PF} and \eqref{eq:norm}, we have 
\begin{equation}
  \mathbb{E}_{\delta_{(x, \Omega)}}[N_n] \sim  k_0^{n}\varphi_0^\dagger(x, \Omega),
  \label{eq:PF2}
\end{equation}
with $\varphi_0^\dagger(x, \Omega)$ given explicitly in Eq.~\eqref{eq:rod_adj_model}. A comparison with respect to Monte Carlo simulations is illustrated in Fig.~\ref{fig:super_N1}, with the same cross sections as in the critical case; the rod half-length $R_0$ has been adjusted according to Eq.~\eqref{eq:rod_L0} in order to ensure $k_0=1.01$. Monte Carlo simulations are again run using $10^6$ particles, with $n$ varying from $n=10$ to $n=100$. The Monte Carlo tally corresponds to the normalized ensemble-averaged number of fission neutrons $\mathbb{E}_{\delta_{(x, \Omega)}}[N_n] / k_0^n$ being in the rod at generation $n$, for a single ancestor neutron starting with coordinates $(x,\Omega)$. The statistical agreement between Monte Carlo results (also normalized by the factor $k_0^n$) and Eq.~\eqref{eq:PF2} is very good. The asymptotic behavior is attained for relatively small values of $n$: at $n=10$, the average number of fission neutrons has already settled to the spatial shape given by Eq.~\eqref{eq:PF2}.

For the second moment, recall from Eq. \eqref{eq:L2sup_mom2} that
\begin{equation}\label{eq:rod-m2}
  \mathbb{E}_{\delta_{(x, \Omega)}}[N_n^2] \sim  k_0^{2n} L_2(x, \Omega) \varphi_0^\dagger(x, \Omega)
\end{equation}
for sufficiently large $n$. Now, note that the combinatorial sum over the set $[\ell_1, \dots, \ell_M]_2^{2+}$ in Eq.~\eqref{eq:L2sup_mom2} contains only one element, $(1, 1)$. Due to this and using the fact that fission is isotropic, the aforementioned combinatorial sum collapses down to $\frac12\mathcal V[\varphi_0^\dagger]$, as given in Eq.~\eqref{eq:calc-V}. It therefore follows that
\begin{equation}
  L_2(x, \Omega) =  \sum_{m = 1}^\infty \frac{\Psi_m^-[\mathcal{V}[\varphi_0^\dagger]](x, \Omega)}{k_0^{2m}\varphi_0^\dagger(x, \Omega)}.  
  \label{eq:PF2mom}
\end{equation}
From a practical point of view, $L_2$ contains terms $\Psi_m^-[\mathcal{V}[\varphi_0^\dagger]](x, \Omega)$, whose analytical expression is known only for large $n$. However, since $\Psi_n^-[\mathcal{V}[\varphi_0^\dagger]](x, \Omega)$ is the expected value of $\mathcal{V}[\varphi_0^\dagger]$ evaluated over the coordinates $(x_i,\Omega_i)$ of neutrons about to undergo a fission event at generation $n \ge 1$ (see Eq.~\eqref{eq:supL}), its expression can be estimated by Monte Carlo simulation. For this purpose, we set
\begin{equation}
    L_2^n(x, \Omega) = \sum_{m = 1}^n \frac{\Psi_m^-[\mathcal{V}[\varphi_0^\dagger]](x, \Omega)}{k_0^{2m}\varphi_0^\dagger(x, \Omega)},
\end{equation}
so that Eq.~\eqref{eq:rod-m2} can written in a more practical form:
\begin{equation}
 \frac{\mathbb{E}_{\delta_{(x, \Omega)}}[N_n^2]}{L_2^n(x, \Omega)} \sim k_0^{2n} \varphi_0^\dagger(x, \Omega)
 \label{eq:rod_super_2}
\end{equation}
for sufficiently large $n$.

\begin{figure}[t]
\centering
\includegraphics[scale = 0.7]{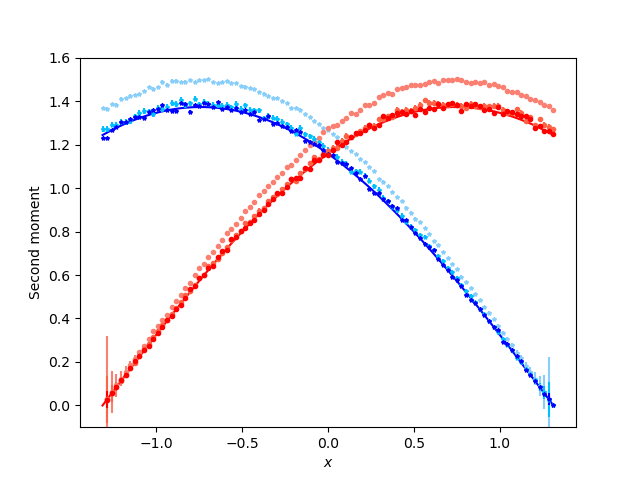}
\caption{Rod model benchmark for supercritical conditions, with $k_0=1.01$: analysis of the second moment of the number of fission neutrons. Comparison between the exact expression for $\varphi_0^\dagger(x, \Omega)$ in Eq.~\eqref{eq:rod_super_2} (solid lines) and Monte Carlo simulations for $\mathbb{E}_{\delta_{(x, \Omega)}}[N_n^2] / (k_0^{2n} L_2^n(x, \Omega) ) $ (symbols). Error bars correspond to one sigma uncertainty. Shades of blue: case $\Omega=+1$, with increasing $n=10$, $n=50$, and $n=100$; shades of red: case $\Omega=-1$, with increasing $n=10$, $n=50$, and $n=100$.}
\label{fig:super_N2}
\end{figure}

A comparison between the analytical result in Eq.~\eqref{eq:rod_super_2} and Monte Carlo simulations is presented in Fig.~\ref{fig:super_N2}, as a function of $n$. The system parameters are the same as for the case of the first moment. Convergence to the asymptotic shape is attained at about $n=50$, which is again later than for the first moment. A good statistical agreement is found between the Monte Carlo findings and the exact formula.

\subsubsection{The subcritical case}

\begin{figure}[t]
\centering
\includegraphics[scale = 0.7]{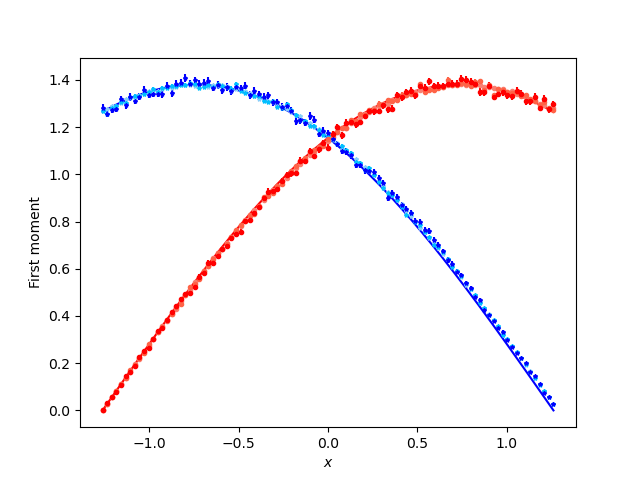}
\caption{Rod model benchmark for subcritical conditions, with $k_0=0.99$: analysis of the average number of fission neutrons. Comparison between the exact expression for $\varphi_0^\dagger(x, \Omega)$ in  Eq.~\eqref{eq:PF2sub} (solid lines) and Monte Carlo simulations for $\mathbb{E}_{\delta_{(x, \Omega)}}[N_n] / k_0^{n}$ (symbols). Error bars correspond to one sigma uncertainty. Shades of blue: case $\Omega=+1$, with increasing $n=10$, $n=50$, and $n=100$; shades of red: case $\Omega=-1$, with increasing $n=10$, $n=50$, and $n=100$.}
\label{fig:sub_N1}
\end{figure}

We finally consider the subcritical case, with $k_0 < 1$. From Eqs.~\eqref{eq:PF} and \eqref{eq:norm}, we have 
\begin{equation}
  \mathbb{E}_{\delta_{(x, \Omega)}}[N_n] \sim k_0^{n}\varphi_0^\dagger(x, \Omega),
  \label{eq:PF2sub}
\end{equation}
where, as usual, $\varphi_0^\dagger(x, \Omega)$ is given explicitly in Eq.~\eqref{eq:rod_adj_model}. A comparison with respect to Monte Carlo simulations is illustrated in Fig.~\ref{fig:sub_N1}, with the same cross sections as in the critical case; the rod half-length $R_0$ has been adjusted according to Eq.~\eqref{eq:rod_L0} in order to ensure $k_0=0.99$. Monte Carlo simulations are again run using $10^6$ particles, with $n$ varying from $n=10$ to $n=100$. The Monte Carlo tally corresponds to the normalized ensemble-averaged number of fission neutrons $\mathbb{E}_{\delta_{(x, \Omega)}}[N_n] / k_0^n$ being in the rod at generation $n$, for a single ancestor neutron starting with coordinates $(x,\Omega)$. The statistical agreement between Monte Carlo results (also normalized by the factor $k_0^n$) and Eq.~\eqref{eq:PF2sub} is very good. Similarly to the supercritical case, the asymptotic behavior is attained for relatively small values of $n$: at $n=10$, the average number of fission neutrons has already settled to the spatial shape given by Eq.~\eqref{eq:PF2sub}.

The second moment stems from Eq.~\eqref{eq:submom2}:
\begin{equation}
    \mathbb{E}_{\delta_{(x, \Omega)}}[N_n^2] \sim  k_0^{n} L_2 \varphi_0^\dagger(x, \Omega).
\end{equation}
Again, noting that the combinatorial sum in the definition of $L_2$ given in \eqref{eq:L2sub_mom2} contains one element, $(1, 1)$, and using the fact that fission is isotropic, yields
\begin{equation}
\label{eq:L2}
L_2 = 1 + 2\Sigma_{\mathtt f} \sum_{m = 0}^\infty \frac{\langle \varphi_0, \mathcal{E}_\cdot[\Psi_m[\mathbf{1}](x, \Omega_1)]^2\rangle}{k_0^{m}},
\end{equation}
where the average $\mathcal{E}$ is taken over the random post-fission directions $\Omega_1$. Similarly to the case of the supercritical configuration, the term $L_2$ contains the expectations $\Psi_m[\mathbf{1}](x, \Omega_1)$, whose analytical expression is only known for large $m$. Again, this term can be estimated by Monte Carlo. To this end, we set
\begin{equation}
 L_2^n = 1 + 2\Sigma_{\mathtt f}\sum_{m = 0}^n \frac{\langle \varphi_0, \mathcal{E}_\cdot[\Psi_m[\mathbf{1}](x, \Omega_1)]^2\rangle}{k_0^{m}}
\end{equation}
for a given fission generation $n$. The inner product appearing in the numerator can be written more explicitly:
\begin{align}
   \langle \varphi_0, \mathcal{E}_\cdot[\Psi_m[\mathbf{1}](x, \Omega_1)]^2\rangle
   &= \frac{1}{4}\int_{-L_0}^{L_0} (\varphi_0(x, +1) + \varphi_0(x, -1))\left(\mathbb E_{\delta_{(x, +1)}}[N_m] + \mathbb E_{\delta_{(x, -1)}}[N_m] \right)^2 dx,
\end{align}
where now the expected values in the integrand can be easily estimated by Monte Carlo and then weighted by the forward eigenfunction given in Eq.~\eqref{eq:rod_for_model}. With this definition, we therefore have 
\begin{equation}
\frac{\mathbb{E}_{\delta_{(x, \Omega)}}[N_n^2]}{L_2^n} \sim k_0^{n} \varphi_0^\dagger(x, \Omega).
\label{eq:rod_sub_2}
\end{equation}

\begin{figure}[t]
\centering
\includegraphics[scale = 0.7]{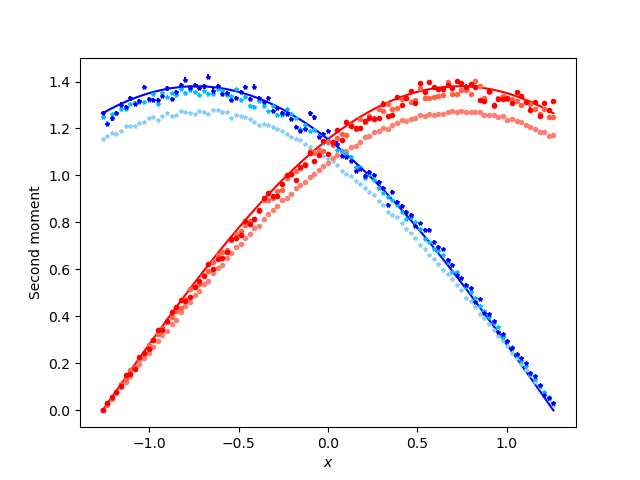}
\caption{Rod model benchmark for subcritical conditions, with $k_0=0.99$: analysis of the second moment of the number of fission neutrons. Comparison between the exact expression for $\varphi_0^\dagger(x, \Omega)$ in Eq.~\eqref{eq:rod_sub_2} (solid lines) and Monte Carlo simulations for $\mathbb{E}_{\delta_{(x, \Omega)}}[N_n^2] / (k_0^{n} L_2^n )$ (symbols). Error bars correspond to one sigma uncertainty. Shades of blue: case $\Omega=+1$, with increasing $n=10$, $n=50$, and $n=100$; shades of red: case $\Omega=-1$, with increasing $n=10$, $n=50$, and $n=100$.}
\label{fig:sub_N2}
\end{figure}

A comparison between the analytical result in Eq.~\eqref{eq:rod_sub_2} and Monte Carlo simulations is presented in Fig.~\ref{fig:sub_N2}, as a function of $n$. The system parameters are the same as for the case of the first moment. Convergence to the asymptotic shape is attained at about $n=50$, which is again later than for the first moment. A good statistical agreement is found between the Monte Carlo findings and the exact formula.

\section{Conclusions}
\label{sec:conclusions}

In this work we have proposed a general formalism to characterize the moments of the neutron population as a function of fission generations, with a particular emphasis on determining reference solutions for Monte Carlo simulations involving adjoint-weighted parameters in $k$-eigenvalue calculations. In particular, precise asymptotics have been established for the average and variance of the neutron population for sub-, super- and critical configurations.

In view of using these findings as a test-bed for the verification of Monte Carlo developments based on the Iterated Fission Probability (IFP) method, which is the most widely adopted strategy to estimate adjoint-weighted quantities in production particle-transport codes, we have introduced a benchmark using the one-dimensional and single-speed `rod-model'. Within this framework, we have determined reference solutions for the asymptotic average number of neutrons stemming from an ancestor particle, as well as for the second moment of this counting process. While the proposed benchmark is necessarily simple in order to allow for analytical solutions, the statements concerning the statistical moments are fairly broad and can be thus extended to more complex configurations.

Future work will concern the generalization of the results proposed in this paper in several directions: exact results for a benchmark configuration using continuous-energy nuclear data might be derived, at least introducing suitable hypotheses on the spatial dependence of the system; furthermore, our strategy might be successfully applied with a limited amount of modifications to the investigation of the generalized IFP method associated to $\alpha$-eigenvalue problems \cite{terranova_ifp}.

\appendix

\section{General statements of the main results}
\label{sec:appendix}

In this section we present the precise statements for the moment asymptotics given informally in Eqs.~\eqref{eq:crit-mom}, \eqref{eq:sup-mom} and \eqref{eq:sub-mom}, as well as the theorem for the survival probability given in Eq.~\eqref{eq:survival} and the Yaglom limit in Eq. \eqref{eq:Yaglom}. The proofs can be found in Ref.~\citenum{NTE-mom-sup}.

Before stating the theorems, we first introduce some assumptions on the model, using the notation defined in Sec. \ref{sec:setting}.
\begin{itemize}
\item[\namedlabel{H1}{(H1)}] The cross sections $\Sigma_\mathtt{s}$, $\Sigma_\mathtt{f}$, $\Sigma_{\mathtt{a}}$ and $\nu_{\mathtt f}$ are uniformly bounded from above. 
\item[\namedlabel{H2}{(H2)}] $\inf \big(\Sigma_\mathtt{f}({\bf r}, E)\chi_{\mathtt f}({\bf r}, {\mathbf \Omega} \to {\mathbf \Omega}', E \to E') + \Sigma_\mathtt{s}({\bf r}, E)f_s({\bf r}, {\mathbf \Omega} \to {\mathbf \Omega}', E \to E')\big) > 0$,
where the infimum is taken over all ${\bf r} \in D$, ${\mathbf \Omega}, {\mathbf \Omega}' \in \mathbb S_2$ and $E, E' \in (E_\mathtt{min}, E_\mathtt{max})$.
\item[\namedlabel{H3}{(H3)}] There exists a constant $C > 0$, such that for any $g \in L_{+, 1}^\infty(\mathcal S)$, the space of non-negative, measurable functions on $\mathcal S$ that are uniformly bounded by unity, we have
\[
  \langle \varphi_0, \Sigma_{\mathtt f}\mathcal V[g]\rangle \ge C\langle \varphi_0, {\color{black}F[g]}\rangle^2, 
\]
where 
\begin{equation}\label{eq:V_proof}
\mathcal{V}[g]({\bf r}, {\mathbf \Omega}, E) 
= \mathcal{E}_{({\bf r}, {\mathbf \Omega}, E)}\left[\left(\sum_{i = 1}^M g({\bf r}, {\mathbf \Omega}_i, E_i)\right)^2 \right] - \mathcal{E}_{({\bf r}, {\mathbf \Omega}, E)}\left[ \sum_{i = 1}^M g^2({\bf r}, {\mathbf \Omega}_i, E_i)\right].
\end{equation}
\end{itemize}

\medskip

Assumption \ref{H1} ensures finite activity in finite time, and \ref{H2} is an irreducibility-type condition that ensures at least one of fission or scatter can occur everywhere in the domain. It was shown in \cite{SNTE3} that under these two assumptions, \eqref{eq:PF} holds. Assumption \ref{H3} can also be thought of as an irreducibility or {\it spread-out-ness} condition at branching events. We note that it is possible to weaken assumption \ref{H3} at the expense of a more complicated proof. However, this is outside the scope of this article and we refer the reader to \cite{Pedro24} for details on how this can be done in continuous time. Finally we note that conditions \ref{H1} and \ref{H2} are clearly satisfied for the rod model presented in Sec. \ref{sec:example}, and we refer the reader to \cite[\S 9]{YaglomNTE} for verification of \ref{H3}.

\renewcommand{\thesubsection}{\thesection.\arabic{subsection}}

\subsection{The critical case}

Suppose $k_0 = 1$. We state three theorems, one each pertaining to the asymptotic behaviour described in Eqs. \eqref{eq:crit-mom}, \eqref{eq:survival} and \eqref{eq:Yaglom}.

\medskip

\begin{theorem}[Critical case, $k_0 = 1$]\label{thm:critnuke} 
 Suppose that \ref{H1} and \ref{H2} hold and $k_0 = 1$. 
Define
 \[
 \Delta_n^{(\ell)} = \sup_{({\bf r}, {\mathbf \Omega}, E) \in \mathcal S, g\in L_{+, 1}^\infty(\mathcal S)}\left| n^{-(\ell-1)} \varphi_0^\dagger({\bf r}, {\mathbf \Omega}, E)^{-1}\Psi_n^{(\ell)}[g]({\bf r}, {\mathbf \Omega}, E)
- 2^{-(\ell-1)} \ell! \,\langle Q_\mathtt{f}, g\rangle^\ell
  \langle\varphi_0, \Sigma_{\mathtt f}\mathcal{V}[\varphi_0^\dagger]\rangle^{\ell-1}\right|,
 \]
where $\mathcal V$ was defined in \eqref{eq:V_proof}.   Then, for all $\ell \ge 1$,
\begin{equation}
\sup_{n\geq0} \Delta_n^{(\ell)}<\infty
\text{ and }
\lim_{n\rightarrow\infty}
\Delta_n^{(\ell)}
=0.
\label{momconv}
\end{equation}
\end{theorem}

\medskip

The next theorem shows that the survival probability behaves asymptotically like $c / n$, for some constant $c > 0$.\

\medskip

\begin{theorem}\label{thm:survival}
Suppose \ref{H1}, \ref{H2} and \ref{H3} hold, and that $k_0 = 1$. Then, for any $({\bf r}, {\mathbf \Omega}, E) \in \mathcal S$, we have
\begin{equation}\label{eq:survival_proof}
\lim_{n \to \infty} n\mathbb{P}_{\delta_{({\bf r}, {\mathbf \Omega}, E)}}(N_n > 0) = \frac{2\varphi_0^\dagger({\bf r}, {\mathbf \Omega}, E)}{\langle \varphi_0, \Sigma_{\mathtt f} \mathcal{V}[\varphi_0^\dagger]\rangle},
\end{equation}
where $\mathcal V$ was defined in \eqref{eq:V_proof}. 
\end{theorem}

\medskip

Finally, we show that the limiting distribution of the process normalised by the current generation converges to an exponential random variable, when conditioned on survival.

\medskip

\begin{theorem}\label{thm:Yaglom}
Suppose \ref{H1}, \ref{H2} and \ref{H3} hold, and that $k_0 = 1$. Then, for any $({\bf r}, {\mathbf \Omega}, E) \in \mathcal S$ and any $f \in L_{+, 1}^\infty(\mathcal S)$, 
\begin{equation}\label{eq:Yaglom_proof}
\lim_{n \to \infty}\left(\frac1n \sum_{i = 1}^{N_n}f({\bf r}_i^{(n)}, {\mathbf \Omega}_i^{(n)}, E_i^{(n)}) \Bigg| N_n > 0\right) = Y,
\end{equation}
where the above convergence is in distribution and $Y$ is an exponential random variable with rate $2/\langle Q_\mathtt{f}, f\rangle \langle \varphi_0, \Sigma_{\mathtt f}\mathcal V[\varphi_0^\dagger]\rangle$.
\end{theorem}

\subsection{Non-critical cases}

We will now consider the moment asymptotics when $k_0 \neq 1$.

\medskip

\begin{theorem}[Supercritical case, $k_0 > 1$]\label{thm:supernuke}
Suppose that \ref{H1} and \ref{H2} hold and $k_0 >1$.  
Define
 \[
 \Delta_n^{(\ell)} = \sup_{({\bf r}, {\mathbf \Omega}, E) \in \mathcal S, g\in L_{+,1}^\infty(\mathcal S)}\left|k_0^{-n\ell} \, \varphi_0^\dagger({\bf r}, {\mathbf \Omega}, E)^{-1}\Psi_n^{(\ell)}[g]({\bf r}, {\mathbf \Omega}, E)
- \langle Q_\mathtt{f}, g \rangle^\ell L_\ell({\bf r}, {\mathbf \Omega}, E)\right|,
 \]
where $L_1 = 1$ and,  for $\ell \ge 2$, $L_\ell({\bf r}, {\mathbf \Omega}, E)$ is given by the recursion
\begin{equation}
L_\ell({\bf r}, {\mathbf \Omega}, E) = \frac{\ell!}{\varphi_0^\dagger({\bf r}, {\mathbf \Omega}, E)} \sum_{m =1}^\infty k_0^{-{\color{black}\ell m}}
\Psi_m^-\Bigg[\mathcal{E}_\cdot\Bigg[\sum_{[\ell_1, \dots, \ell_M]_\ell^{2+}}\prod_{\substack{j = 1 \\ \ell_j > 0}}^M L_{\ell_j}(\cdot, {\mathbf \Omega}_j, E_j)\varphi_0^\dagger(\cdot, {\mathbf \Omega}_j, E_j)\Bigg]\Bigg]({\bf r}, {\mathbf \Omega}, E),
\label{supL}
\end{equation}
with $[\ell_1, \dots, \ell_M]_\ell^{2+}$ defining the set of non-negative tuples $(\ell_1, \dots, \ell_N)$, such that $\sum_{j = 1}^N \ell_j = \ell$ and at least two of the $\ell_j$ are strictly positive.  

Then, for all $\ell\ge 1$,
\begin{equation*}
\sup_{n\geq 0} \Delta_n^{(\ell)}<\infty
\text{ and }
\lim_{n\rightarrow\infty}
\Delta_n^{(\ell)}
=0.
\end{equation*}
\end{theorem}

\medskip

Finally, we consider the subcritical case.

\medskip

\begin{theorem}[Subcritical case, $k_0 < 1$]\label{thm:subnuke}
Suppose that \ref{H1} and \ref{H2} hold and $k_0 < 1$.  
Define
 \[
 \Delta_n^{(\ell)} = \sup_{({\bf r}, {\mathbf \Omega}, E) \in \mathcal S, g\in L_{+,1}^\infty(\mathcal S)}\left|
 k_0^{-n}\, \varphi_0^\dagger({\bf r}, {\mathbf \Omega}, E)^{-1}\Psi_n^{(\ell)}[g]({\bf r}, {\mathbf \Omega}, E)
-  L_\ell\right|,
 \]
where $L_1 = \langle Q_\mathtt{f}, g\rangle$ and for $\ell \ge 2$, $L_\ell$ is given by 
\[
L_\ell =  \langle Q_\mathtt{f},  g^\ell \rangle + \sum_{m = 0}^\infty k_0^{-m}\left\langle \varphi_0 , \Sigma_{\mathtt f}(\cdot)\mathcal{E}_\cdot\Bigg[\sum_{[\ell_1, \dots, \ell_N]_\ell^{2+}}{\ell \choose \ell_1 \dots \ell_N}\prod_{{\substack{j = 1 \\ \ell_j > 0}}}^N\Psi_m^{(\ell_j)}[f](\cdot, {\mathbf \Omega}_j, E_j)\Bigg]\right\rangle,
\]
where $[\ell_1, \dots, \ell_N]_\ell^{2+}$ is as in the previous theorem. 

Then, for all $\ell \ge 1$,
\begin{equation*}
\sup_{n\geq 0} \Delta_n^{(\ell)}<\infty
\text{ and }
\lim_{n\rightarrow\infty}
\Delta_n^{(\ell)}
=0.
\end{equation*}
\end{theorem}

\medskip

\section*{Acknowledgments}

This work was partially supported by the EPSRC grant EP/W026899/1. The authors report there are no competing interests to declare.

\bibliography{bibliography}
\bibliographystyle{plain}

\end{document}


\maketitle
\begin{abstract}
We show that the results presented in \cite{NTE-mom} actually hold for a much broader class of discrete time branching processes. 
\end{abstract}

\section{Introduction}
Consider a discrete time spatial branching process, $(\mathcal{X}_n)_{n \ge 0}$, on a locally compact Hausdorff space, $U$, where $\mathcal{X}_n$ denotes the configuration of particles in the $n$-th generation of the population. 
At each unit of time, a branching event occurs such that, if the parent is at $x \in U$, the new configuration of particles is given by the point process $(\mathcal{Z}, \tilde\cP_x)$, where
\[
  \cZ = \sum_{i = 1}^N \delta_{z_i}.
\]
In this setting, we also allow for the possibility of absorption by allowing $\tilde\cP_x(N = 0) > 0$. 
\smallskip

Letting $N_n$ denote the number of individuals in the $n^{th}$ generation, the generational branching process is formally defined via the collection of atomic measures
\[
  \mathcal X_n = \sum_{i = 1}^{N_n}\delta_{x_i^{(n)}}, \qquad n \ge 0,
\]
where $\{x_i^{(n)} : i = 1, \dots, N_n\}$ denotes the collection of particles in the $n$-th generation. Setting $M(U)$ to be the set of finite counting measures on $U$, we let $(\mathbb{P}_\mu, \mu \in M(U))$ denote the law of the branching process defined above.

\smallskip

The associated linear semigroup is given by
\begin{equation}
  \Psi_n[g](x) := \mathbb{E}_{\delta_x}\Bigg[ \sum_{i = 1}^{N_n}g(x_i^{(n)})\Bigg], \qquad n \ge 1, \, x \in U, \, g \in B(U),
  \label{lin-semi}
\end{equation}
with $\Psi_0[g](x) = g(x)$ and where we have used $B(U)$ to denote the set of non-negative uniformly bounded measurable functions on $U$. Moreover, if we define the linear branching mechanism
\begin{equation}
  \sF[g](x) = \tilde{\mathcal{E}}_x\Bigg[\sum_{i = 1}^Ng(z_i)\Bigg], \qquad x \in U, \, g \in B(U),
  \label{bmechlin}
\end{equation}
the linear evolution equation associated with $(\Psi_n, n \ge 0)$ is given by
\begin{equation}
\Psi_n[g](x) =   \sF[\Psi_{n-1}[g]](x).
\label{linrecursion}
\end{equation}

\bigskip

It is worth noting that we can think of such processes as being embedded into a continuous-time branching Markov process for which, in between generations, particles move according to a (sub)Markov process $(\xi, \mathbf{P})$. This is precisely the setting described in \cite{NTE-mom}. Indeed, particles move piecewise deterministically between scattering events until either the particle is absorbed or there is a fission event, contributing to the next generation of particles in $\mathcal{X}$. Note, however, that there is a slight difference in how the offspring distribution is described: in the general setting, $\mathcal{Z}$ describes the positions of the offspring relative to the position of the parent particle at its birth time, whereas in the NTE, the configurations of the offspring were given relative to the configuration of the particle at the time of branching. Of course, with a little work, these two perspectives can be reconciled.


\bigskip

As in \cite{NTE-mom}, we will study the moment operators 
\[
\Psi_n^{(k)}[g](x) :=  \mathbb{E}_{\delta_x}\Bigg[\left( \sum_{i = 1}^{N_n}g(x_i^{(n)})\right)^k\Bigg], \qquad n \ge 1, \, x \in U, \, g \in B(U),
\]
and show that Theorems 1, 4 and 5 from \cite{NTE-mom} hold under appropriate assumptions on $\mathcal{X}$. In the critical case, we will then show that the asymptotic for the survival probability combined with the moment asymptotics yields the Yaglom limit result. We now introduce some relevant assumptions, {\color{black}using the notation $\langle \eta, f\rangle = \int_U f(x)\eta(\d x)$, when $\eta$ is a measure on $U$ and $f \in B(U)$.}

\bigskip

\begin{itemize}
\item[\namedlabel{G1}{(G1)}] There exists a constant $\rho > 0$, a function $\varphi \in B(U)$ and finite measure $\eta\in M(U)$ such that, for $g\in B(U)$ and $\mu\in M(U)$, 
\begin{equation}\label{eq:eigentriple}
\langle\eta, \Psi_n[g]\rangle = \rho^n\langle \eta, g\rangle \quad \text{ and } \quad \langle\mu, \Psi_n[\varphi]\rangle
=\rho^n\langle \mu,\varphi\rangle, \quad n\geq0.
\end{equation}
Moreover, defining
\begin{equation}
 \Delta_n = \sup_{x\in E, g\in B(U)}|\varphi(x)^{-1}\rho^{-n}\Psi_n[f](x)-\langle \eta, f \rangle| , \qquad t\geq 0,
\label{Deltan}
 \end{equation}
we have
\begin{equation}
\sup_{n\geq 1}\Delta_n <\infty \text{ and }\lim_{n\to\infty} \Delta_n=0. 
\label{T1ass}
\end{equation}

\item[\namedlabel{G2_k}{(G2)$_k$}]
Suppose $k \ge 1$. Then
\begin{equation}
  \sup_{x\in E}\tilde{\mathcal{E}}_x(\langle 1, \mathcal{Z}\rangle^k) <\infty.
\label{offspring-mom}
\end{equation}

\item[\namedlabel{G3}{(G3)}]
There exists $n_{\mathtt{max}} \ge 1$ such that $\mathcal{P}_x(N \le n_{\mathtt{max}}) = 1$ for all $x \in U$.

\item[\namedlabel{G4}{(G4)}] 
There exists a constant $C > 0$, such that
\[
  \langle \eta, \tilde{\mathcal{V}}[g]\rangle \ge C\langle \eta, g\rangle^2, 
\]
where, for $g \in B(U)$, 
\begin{equation}\label{eq:V}
\tilde{\mathcal{V}}[g](x) 
= \tilde{\mathcal{E}}_{x}\Bigg[\left(\sum_{i = 1}^N g(z_i)\right)^2 \Bigg] - \tilde{\mathcal{E}}_{x}\Bigg[ \sum_{i = 1}^N g^2(z_i)\Bigg].
\end{equation}

\item[\namedlabel{G5}{(G5)}] 
For all $n$ sufficiently large, $\sup_{x \in U} \mathbb P_{\delta_{x}}(N_n > 0) < 1$.
\end{itemize}

\bigskip

The reader will note that the first assumption is the analogue, albeit stated more precisely, of equation (11) in \cite{NTE-mom}, for which (H1) and (H2) are sufficient conditions, see Ref. \cite{SNTE3}. 
Note that \eqref{eq:eigentriple} implies that $\rho^n$ is the leading eigenvalue for the semigroup $\Psi_n$ with corresponding right eigenfunction $\varphi$ and left eigenmeasure $\eta$. In the same way that $k_{\rm eff}$ describes the criticality of the NTE, $\rho$ categorises the criticality of the branching process. The reader will also notice that $\eta$ is a measure in this case. For assumptions \ref{G2_k} and \ref{G3}, these provide some control over the offspring distribution. In the setting of the NTE, these assumptions are not necessary since the number of neutrons produced at fission events is almost surely bounded. Assumption \ref{G4} is identical to that of the NTE setting. Finally, assumption \ref{G5} ensures that there are no points in $U$ from which the process could survive almost surely. This final assumption is not needed in the case of the NTE since it holds due to the fact that the spatial domain $D$ is bounded and all energies are bounded below by $E_{\mathtt{min}} > 0$, which means all particles have a velocity that is bounded below by say $v_{\mathtt{min}} > 0$.

\section{Main results}
\subsection{Critical case}
We first focus on the critical case and hence assume that $\rho = 1$. We present the analogues of Theorems 1, 2 and 3 of \cite{NTE-mom}.

\begin{theorem}[Critical, $\rho = 1$]\label{thm:crit} Suppose that \ref{G1} holds along with \ref{G2_k} for some $k \ge 2$ and $\rho = 1$. 
Define
 \[
 \Delta_n^{(\ell)} = \sup_{x\in E, g\in B(U)}\Bigg|n^{-(\ell-1)} \varphi(x)^{-1}\Psi_n^{(\ell)}[g](x)
- 2^{-(\ell-1)} \ell! \,\langle \eta, f \rangle^\ell
\langle  \eta, \tilde{\mathcal{V}}[\varphi]\rangle^{\ell-1}\Bigg|,
 \]
where 
\[
\tilde{\mathcal{V}}[\varphi](x) 
=\tilde{\mathcal{E}}_{x}\Bigg[\langle\varphi, \mathcal{Z}\rangle^2 - \langle \varphi^2, \mathcal{Z}\rangle\Bigg]
= \tilde{\mathcal{E}}_{x}\Bigg[ \sum_{i = 1}^N \sum_{\substack{j = 1 \\ j\neq i}}^N \varphi(z_i)\varphi(z_j)\Bigg]
\]

Then, for all $\ell\leq k$
\begin{equation}
\sup_{n\geq1} \Delta_n^{(\ell)}<\infty
\text{ and }
\lim_{n\rightarrow\infty}
\Delta_n^{(\ell)}
=0.
\label{momconv}
\end{equation}
\end{theorem}

\bigskip

\begin{theorem}\label{thm:survival}
Suppose \ref{G1}, \ref{G3}, \ref{G4} and \ref{G5} hold. Then, for any $x \in U$, we have
\begin{equation}\label{eq:survival}
\lim_{n \to \infty} n\mathbb{P}_{\delta_{x}}(N_n > 0) = \frac{2\varphi(x)}{\langle \eta, \tilde{\mathcal{V}}[\varphi]\rangle},
\end{equation}
where $\mathcal V$ was defined in \eqref{eq:V}.
\end{theorem}

\bigskip

\begin{theorem}\label{thm:Yaglom}
Suppose \ref{G1}, \ref{G3}, \ref{G4} and \ref{G5} hold. Then, for any $x \in U$ and any $f \in B(U)$, 
\begin{equation}\label{eq:Yaglom}
\lim_{n \to \infty}{\text Law}\left(\frac1n \sum_{i = 1}^{N_n}f(x_i^{(n)}) \Bigg| N_n > 0\right) =^d Y,
\end{equation}
where the above convergence is in distribution and $Y$ is an exponential random variable with rate $2/\langle \eta, f\rangle \langle \eta, \tilde{\mathcal{V}}[\varphi]\rangle$.
\end{theorem}

%

\subsection{Non-critical cases}
Now we give the analogues of Theorems 4 and 5 in \cite{NTE-mom}.
\begin{theorem}[Supercritical, $\rho > 1$]\label{thm:super}
 Suppose that \ref{G1} holds along with \ref{G2_k} for some $k \ge 2$ and $\rho > 1$. 
Define
 \[
 \Delta_n^{(\ell)} = \sup_{x\in E, g\in B(U)}\Bigg|\rho^{-n\ell}\varphi(x)^{-1}\Psi_n^{(\ell)}[g](x)
- {\ell!} \,\langle \eta, f \rangle^\ell L_\ell(x)\Bigg|,
 \]
where $L_1 = 1$ and for $k \ge 2$, $L_k(x)$ is given by the recursion
\begin{equation}
L_k(x) = \varphi(x)^{-1}\sum_{n =0}^\infty \rho^{-{k(n+1)}} \Psi_n\Bigg[\tilde{\mathcal{E}}_\cdot\Bigg[\sum_{[k_1, \dots, k_N]_k^{2+}}\prod_{\substack{j = 1 \\ k_j > 0}}^N L_{k_j}(z_j) \varphi(z_j)\Bigg]\Bigg](x)
\label{supL}
\end{equation}
with $[k_1, \dots, k_N]_k^{2+}$ defining the set of non-negative tuples $(k_1, \dots, k_N)$, such that $\sum_{j = 1}^N k_j = N$ and at least two of the $k_j$ are strictly positive.  

Then, for all $\ell\leq k$
\begin{equation*}
\sup_{n\geq 1} \Delta_n^{(\ell)}<\infty
\text{ and }
\lim_{n\rightarrow\infty}
\Delta_n^{(\ell)}
=0.
\end{equation*}
\end{theorem}

\bigskip

\begin{theorem}[Subcritical, $\rho < 1$]\label{thm:sub}
 Suppose that \ref{G1} holds along with \ref{G2_k} for some $k \ge 2$ and $\rho < 1$. 
Define
 \[
 \Delta_n^{(\ell)} = \sup_{x\in E, g\in B(U)}\Bigg|
 \rho^{-n} \varphi(x)^{-1}\Psi_n^{(\ell)}[g](x)
-  L_\ell\Bigg|,
 \]
where $L_1 = \langle \eta, g \rangle$ and for $\ell \ge 2$, $L_\ell$ is given by
\[
L_\ell =  \langle  \eta, g^\ell\rangle  + \sum_{n = 0}^\infty \rho^{-(n+1)} \left\langle \eta, \tilde{\mathcal{E}}_\cdot\Bigg[\sum_{[k_1, \dots, k_N]_k^{2+}}{k \choose k_1,\cdots, k_N}\prod_{\substack{j = 1 \\ k_j > 0}}^N\Psi_n^{(k_j)}[g](z_j)\Bigg]\right\rangle.
\]

Then, for all $\ell\leq k$
\begin{equation*}
\sup_{n\geq 1} \Delta_n^{(\ell)}<\infty
\text{ and }
\lim_{n\rightarrow\infty}
\Delta_n^{(\ell)}
=0.
\end{equation*}
\end{theorem}

\section{Proofs}\label{sec:proofs}
Recall the recursion \eqref{linrecursion}. In order to prove Theorem 4, we will also need to introduce the corresponding non-linear recursion. Define the non-linear semigroup
\begin{equation}
  \Phi_n[g](x) := \mathbb{E}_{\delta_x}\Bigg[ \prod_{i = 1}^{N_n}g(x_i^{(n)})\Bigg], \qquad n \ge 1, \, x \in U, \, g \in B(U),
  \label{non-lin-semi}
\end{equation}
with $\Phi_0[g] (x)= g(x)$ and where $B(U)$ denotes the set of non-negative measurable functions on $U$ that are bounded by unity.  In addition, we define the non-linear branching mechanism via
\begin{equation}
  \sG[g](x) = \mathcal{E}_x\Bigg[\prod_{i = 1}^Ng(z_i)\Bigg] , \qquad x \in U, \, g \in B(U).
  \label{bmech}
\end{equation}
Analogously to \eqref{linrecursion}, we have the non-linear evolution equation
\begin{equation}
\Phi_n[g](x) =   \sG[\Phi_{n-1}[g]](x), \qquad n\geq 1.
\label{nonlinrecursion}
\end{equation}

The aim is to develop the above recursions to give a more convenient evolution equation involving both $\Psi_n$ and $\Phi_n$, which will help us to more easily obtain expressions for the moments $\Psi_n^{(k)}$. Starting with \eqref{nonlinrecursion} and noting that from \eqref{linrecursion}, we have $\Psi_1[g](x)= \sF[g](x)$, we can add and subtract terms to yield
\[
\Phi_n[g](x) =   (\sG-\sF)[\Phi_{n-1}[g]](x)+ \Psi_1[\Phi_{n-1}[g]](x), \qquad n\geq 2, x\in E, g\in B(U).
\]
Using the same trick with the second term on the right-hand side of the above representation, we have 
\begin{align*}
\Phi_n[g](x) &= (\sG-\sF)[\Phi_{n-1}[g]](x)+\Psi_1[(\sG-\sF)[\Phi_{n-2}[g]]](x) + \Psi_1[\Psi_1[\Phi_{n-2}[g]]](x)\\
&=(\sG-\sF)[\Phi_{n-1}[g]](x)+\Psi_1[(\sG-\sF)[\Phi_{n-2}[g]]](x) +\Psi_2[\Phi_{n-2}[g]](x).
\end{align*}
where we have used the fact that $\Psi_n$ is a semigroup and thus
\[
\Psi_n[g] (x)= \underbrace{\Psi_1[\Psi_1[\cdots\Psi_1}_{n \text{ times }}[g]]](x).
\]
Continuing this recursion, we obtain
\begin{equation}
\Phi_n[g](x) = \sum_{\ell = 0}^{n-1} \Psi_\ell[(\sG-\sF)[\Phi_{n-\ell-1}[g]]](x) + \Psi_n[g](x), \qquad n\geq 2, x\in E, g\in B(U).
\label{keyrecursion}
\end{equation}
Now, it is a straightforward calculation to see that 
\begin{equation}
\Psi^{(k)}_n[f](x) = (-1)^{k}\frac{\partial^k}{\partial\theta^k}\Phi_n[{\rm e}^{-\theta f}](x)\Bigg|_{\theta = 0}\qquad n\geq 2, x\in E, f \in B(U).
\label{diff}
\end{equation}
Moreover, in the spirit of Proposition 1 of \cite{moments}, the following proposition is straightforward to develop appealing directly to \eqref{keyrecursion} and \eqref{diff}. We omit its proof for the sake of brevity.
\begin{prop}
\label{prop}
For $n\geq 1$, $x\in E$ and $f\in B(U)$, we have
\begin{equation}\label{kmomentcomplex}
 \Psi^{(k)}_n[f](x) = \Psi_n[f^k](x) + \sum_{\ell = 0}^{n-1} \Psi_\ell\Bigg[
 \tilde{\mathcal{E}}_\cdot\Bigg[
 \sum_{[k_1,\cdots, k_N]^{2+}_k}{k \choose k_1, \dots, k_N}\prod_{j = 1}^N \Psi^{(k_j)}_{n-\ell-1}[f](z_j) 
 \Bigg]
 \Bigg](x),
\end{equation} 
where $[k_1,\cdots, k_N]^{2+}_k$ is the set of all non-negative $N$-tuples $(k_1, \dots, k_N)$ such that $\sum_{i = 1}^N k_i = k$ and at least two of the $k_i$ are strictly positive.
\end{prop}

With \eqref{kmomentcomplex} in hand, we are ready to move forward to the proofs of the main results, which we split into three parts, depending on whether $\rho > 1$, $< 1$ or $= 1$.

\subsection{Critical case, $\rho = 1$} 
\subsubsection*{Proof of Theorem \ref{thm:crit}}
Following the reasoning in \cite{moments}, we argue via induction. To this end, let us assume that the conclusion of Theorem \ref{thm:crit} holds for all moments of order $\ell\leq k-1$.

First note that we can discard the term $ \Psi_n[f^k](x)$ in \eqref{kmomentcomplex}. Indeed, since $f\in B(U)$, and $\Psi_n[f^k] \sim \langle \eta, f^k\rangle\varphi$ uniformly as $n\to\infty$, we have 
\[
\lim_{n\to\infty}n^{-(k-1)} \Psi_n[f^k](x)  = 0.
\]

Now denote by  $[k_1,\cdots, k_N]^{3+}_k$, the subset of tuples in  $[k_1,\cdots, k_N]^{2+}_k$ for which at least three of the $k_i$ are positive. Turning to the second term on the right-hand side of \eqref{kmomentcomplex}, we have
\begin{align}
&\lim_{n\to\infty}n^{-(k-1)}\sum_{\ell = 0}^{n-1} \Psi_\ell\Bigg[
 \tilde{\mathcal{E}}_\cdot\Bigg[
 \sum_{[k_1,\cdots, k_N]^{3+}_k}{k \choose k_1, \dots, k_N}\prod_{j = 1}^N \Psi^{(k_j)}_{n-\ell-1}[f](z_j) 
 \Bigg]
 \Bigg](x)\notag\\
&= \lim_{n\to\infty}\frac{1}{n}\sum_{\ell = 0}^{n-1} \Psi_\ell\Bigg[
 \tilde{\mathcal{E}}_\cdot\Bigg[
 \sum_{[k_1,\cdots, k_N]^{3+}_k}{k \choose k_1, \dots, k_N}n^{-(\#\{j:k_j>0\}-2)}\prod_{j = 1}^N n^{-(k_j-1)}\Psi^{(k_j)}_{n-\ell-1}[f](z_j)
 \Bigg]
 \Bigg](x)
 \label{biggerthan3}
\end{align}
Under the hypothesis \ref{G1} and since  $\#\{j:k_j>0\}\geq 3$, the limit as $n\to\infty$ of the scaled product that appears in the summands on the right-hand side above tends uniformly to zero.
Still appealing to \ref{G1}, suppose we fix $\varepsilon>0$ and accordingly choose $n_0\in\mathbb{N}$ such that $\Delta_n^{(1)}$ defined in the statement of the theorem is smaller than $\varepsilon$ for all $n>n_0$. From the factor $1/n$ preceding the sum on the right-hand side of \eqref{biggerthan3}, the uniformly diminishing scaled product in the summands and the observation that under \ref{G2_k} 
\begin{equation}
 \sum_{[k_1,\cdots, k_N]^{2+}_k}{k \choose k_1, \dots, k_N}\leq N^k,
\label{DCT}
\end{equation}
it is straightforward to show that the limit of the first $n_0$ terms in the sum on the right-hand side of \eqref{biggerthan3} will tend to zero.
On the other hand, for the part of the sum between $n_0+1$ and $n-1$, the reasons just stated, coupled with uniformly close to limiting behaviour of $\Psi_\ell$ and that the factor $1/n$ on the right-hand side of \eqref{biggerthan3} compensates the growing number of elements in the sum ensures that the remaining part of the limit on the right-hand side of \eqref{biggerthan3} also tends to zero.

To deal with the limit of \eqref{kmomentcomplex}, it thus suffices to replace the summation on the right-hand side  over only pairs  $k_1, k_2>0$ such that $k_1+k_2 = k$ (being careful not to double count such pairs). Our task is thus to show that  
\begin{align*}
\Delta^{(k),2}_{n}=\sup_{x\in E, f\in B(U)}\Bigg|\frac{1}{n \varphi}\sum_{\ell = 0}^{n-1} \left( \frac{n-\ell-1}{n}\right)^{k-2} \Psi_\ell[
 \sH_{n-\ell-1}^{(k)}[f] ]-2^{-(k-1)} k! \,\langle \tilde\varphi, f \rangle^k
\langle  \eta, \tilde{\mathcal{V}}[\varphi]\rangle^{k-1} \Bigg|
\end{align*}
tends to zero as $n\to\infty$, where 
\[
\sH_{m}^{(k)}[f](x)= \frac{1}{2}\tilde{\mathcal{E}}_x\Bigg[
 \sum_{i = 1}^N\varphi(z_i)\sum_{\substack{j = 1\\ j\neq i}}^N\varphi(z_j)
 \sum_{k_1 = 1}^{k-1}{k \choose k_1 }\frac{\Psi^{(k_1)}_{m}[f](z_i)}{\varphi(z_i) m^{k_1-1}} 
 \frac{\Psi^{(k-k_1)}_{m}[f](z_j)}{\varphi(z_j) m^{k-k_1-1}} \Bigg].
\]

Thanks to \eqref{DCT}, the boundedness of $\varphi$, (H2)$_k$ and the induction hypothesis, we see that, for fixed $k$, 
\begin{align}
\sH_{\infty}^{(k)}[f](x): = \lim_{m\to\infty}\sH_{m}^{(k)}[f](x) 
&=(k-1)k!\tilde{\mathcal{V}}[\varphi](x) 2^{-(k-1)}\langle  \eta, \tilde{\mathcal{V}}[\varphi]\rangle^{k-2}
\langle \tilde\varphi, f \rangle^{k},
\label{impliesAs}
\end{align}
where in fact, the convergence can be taken uniformly in both $x\in E$ and $f\in B(U)$.

Next, we want to apply  Theorem \ref{ergodic} in the Appendix. In order to do so,  we can identify $$\varphi(x)F[f](x, \ell, n) =   \left(\frac{n-\ell-1 }{n}\right)^{k-2}\sH_{n-\ell-1}^{(k)}[f](x), $$
in which case, for fixed $\ell\leq n$,
\begin{equation}
\varphi(x)\hat{F}[f](x, \ell,n) = \left(\frac{n-\ell-1}{n} \right)^{k-2} \sH_\infty^{(k)}[f](x).
\label{hatF}
\end{equation}
The verification of \eqref{A2} is relatively straightforward with 
\begin{align*}
\varphi(x)\check{{F}}[f](x) &=\sH_\infty^{(k)}[f](x) \lim_{n\to\infty}\frac{1}{n^{k-1}}\sum_{\ell = 0}^{n-1}  ( n-\ell-1 )^{k-2}\notag\\
&=\sH_\infty[f](x)
 \lim_{n\to\infty}\frac{(n-n_0)^{k-1}}{(k-1)n^{k-1}}\notag\\
 &=\frac{\sH_\infty[f](x)}{(k-1)}
\end{align*}
Note, the above calculation follows from Faulhaber's formula, or equivalently Bernoulli's {\it Summae Potestatum}, and, to leading order, behaves like its integral analogue in the limit. 

In order to deal with \eqref{triangle}, we note that
\begin{align*}
  &\lim_{n\to\infty} \sup_{   f\in B(U)}
\Bigg|
\frac{1}{n
}\sum_{\ell = 0}^{n-1} \langle   \eta, \varphi \hat{F}[f](\cdot, \ell, n) \rangle 
-\langle \eta ,\varphi \check{{F}}[f]\rangle 
\Bigg|\notag\\
&=\frac{\langle \eta, \sH[f]\rangle}{(k-1)}  \lim_{n\to\infty} \sup_{   f\in B(U)} \langle\eta, g\rangle^k
\Bigg|
\frac{1}{n^{k-1}}\sum_{\ell = 0}^{n-1}(k-1) ( n-\ell-1 )^{k-2}
-1
\Bigg|\notag\\
&=0,
\end{align*}
where again, we have used the convergence of Faulhaber's series and the fact that $ \langle\eta, f\rangle^k$ is uniformly bounded for $f\in B(U)$.

Inspecting \eqref{hatF}, assumption \eqref{ass1} is trivially satisfied thanks to the uniform boundedness of $\langle \tilde\varphi, f \rangle^{k}$ for $f\in B(U)$ as well as the uniform boundedness of $\tilde{\mathcal{V}}[\varphi](x) $ thanks to the assumption (H2)$_k$.

Finally, for \eqref{ass2} we have
\begin{align*}
\varphi(x)| F[f](x,\ell, n)-\hat{F}[f](x,\ell, n)|= \left(\frac{n-\ell-1}{n} \right)^{k-2} |\sH_{n-\ell-1}[f](x)- \sH_\infty[f](x)|.
\end{align*}
Hence \eqref{ass2} is a consequence of the uniformity in the convergence \eqref{impliesAs} when $(n-\ell-1)/n$ is bounded away from any small $\varepsilon$. Otherwise, when the pre-factor $((n-\ell-1)/n)^{k-2}$ is arbitrarily small, thanks to the assumption (H2)$_k$ and the induction hypothesis, which ensures that 
\[
\sup_{x\in E, m\geq0, f\in B(U)}\sH_m[f](x)<\infty,
\]
 \eqref{ass2} also holds. 

 It now follows from Theorem \ref{ergodic} that 
 \[
 \sup_{n\geq 2}\Delta^{(k),2}_{n}<\infty \text{ and }\lim_{n\to\infty}\Delta^{(k),2}_{n} = 0.
 \]
 This completes the proof.

\subsubsection*{Proofs of Theorems \ref{thm:survival} and \ref{thm:Yaglom}}
The proof of Theorem \ref{thm:survival} follows the same steps as the continuous time case, \cite{yaglomNTE}. We leave the details as an exercise to the reader. Theorem \ref{thm:Yaglom} now follows from Theorems \ref{thm:crit} and \ref{thm:survival} and the following observation.
\[
\frac{1}{n^k}\mathbb{E}_{\delta_x}[\langle f, \mathcal X_n\rangle^k | N_n > 0]
= \frac{\frac{1}{n^{k-1}}\mathbb{E}_{\delta_x}[\langle f, \mathcal X_n\rangle^k]}{n\mathbb P_{\delta_x}(N_n > 0)}.
\]

\subsection{Supercritical case, $\rho > 1$} 
Let us assume for induction that the statement of the theorem holds for all moments less than or equal to $k-1$.
As with the critical setting, we can  ignore the first term on the right-hand side of \eqref{kmomentcomplex} when scaling the $k$-th moment, this time, by $\rho^{-kn}$. 
Organising the second term on the right-hand side of \eqref{kmomentcomplex}, our objective is to address the limit of
 \begin{align}
& 
\Big| \rho^{-k n}\varphi(x)^{-1}\Psi^{(k)}_n[f](x)- k!\langle \eta, f\rangle^k L_k(x)\Big|\notag\\
&=  
\varphi(x)^{-1}\Bigg|  \sum_{\ell = 0}^{n-1} \rho^{-k(\ell + 1)}\Psi_\ell\Bigg[
k!  \tilde{\mathcal{E}}_\cdot\Bigg[
 \sum_{[k_1,\cdots, k_N]^{2+}_k}\prod_{\substack{j = 1 \\ k_j > 0}}^N \frac{\rho^{-k_j (n-\ell-1)}\Psi^{(k_j)}_{n-\ell-1}[f](z_j)}{k_j!\varphi(z_j)}\varphi(z_j)
 \Bigg]
 \Bigg]\notag\\
 &\hspace{3cm}- k!\langle\eta, f\rangle^k \sum_{\ell =0}^\infty \rho^{-k(\ell + 1)}\Psi_\ell\Bigg[\tilde{\mathcal{E}}_\cdot\Bigg[\sum_{[k_1, \dots, k_N]_k^{2+}}\prod_{\substack{j = 1 \\ k_j > 0}}^N L_{k_j}(z_j)\varphi(z_j) \Bigg]\Bigg](x)\Bigg|,
 \label{truncate}
 \end{align}
where we recall that
\begin{equation}
L_k(x) = \varphi(x)^{-1}\sum_{n =0}^\infty \rho^{-k(n+1)} \Psi_n\Bigg[\tilde{\mathcal{E}}_\cdot\Bigg[\sum_{[k_1, \dots, k_N]_k^{2+}}\prod_{\substack{j = 1 \\ k_j > 0}}^N L_{k_j}(z_j)\varphi(z_j) \Bigg]\Bigg](x).
\label{supL}
\end{equation}
It is easy to see that the convergence of the sum in  \eqref{supL} is uniform in $x\in E$ and hence we can replace the infinite sum on the right-hand side of \eqref{truncate} with the sum from $0$ to $n-1$, since the tail of the sum converges to zero. 
Thus, we may rewrite the right-hand side of the equality in \eqref{truncate} as
\begin{align}
\Theta_n&:=\varphi(x)^{-1}k! \sum_{\ell = 0}^{n-1} \rho^{-k(\ell + 1)}\Psi_\ell\Bigg[\tilde{\mathcal{E}}_\cdot \Bigg[\sum_{[k_1, \dots, k_N]_k^{2+}}\Bigg|\prod_{\substack{j = 1 \\ k_j > 0}}^N \frac{\rho^{-k_j (n-\ell-1)}\Psi^{(k_j)}_{n-\ell-1}[f](z_j)}{k_j! \varphi(z_j)}\varphi(z_j) \notag\\
&\hspace{8cm}
- \prod_{\substack{j = 1 \\ k_j > 0}}^N \langle\eta, f\rangle^{k_j}L_{k_j}(z_j)\varphi(z_j)\Bigg|\Bigg]\Bigg](x).
\label{tendsto0}
\end{align}
It thus remains to show that $\lim_{n \to \infty}\Theta_n = 0$. We first deal with the `lower end' of the sum in \eqref{tendsto0}. To this end, fix $n_0 \ge 1$ and $\varepsilon > 0$ and note that since $|\prod a_i - \prod b_i|\leq \sum|a_i-b_i|$, for positive $a_i,b_i$, we have
\begin{align}
\sum_{\ell = 0}^{n_0} &\rho^{-k(\ell + 1)}\Psi_\ell\Bigg[\tilde{\mathcal{E}}_\cdot \Bigg[\sum_{[k_1, \dots, k_N]_k^{2+}}\Bigg|\varphi(x)^{-1}k!\prod_{\substack{j = 1 \\ k_j > 0}}^N \frac{\rho^{-k_j (n-\ell-1)}\Psi^{(k_j)}_{n-\ell-1}[f](z_j)}{k_j! \varphi(z_j)}\varphi(z_j)
- \prod_{\substack{j = 1 \\ k_j > 0}}^N \langle\eta, f\rangle^{k_j}L_{k_j}(z_j)\varphi(z_j)\Bigg|\Bigg]\Bigg](x)\notag\\
&\le \varphi(x)^{-1}k!\sum_{\ell = 0}^{n_0} \rho^{-k(\ell + 1)}\Psi_\ell\Bigg[\tilde{\mathcal{E}}_\cdot \Bigg[\sum_{[k_1, \dots, k_N]_k^{2+}}\sum_{\substack{j = 1 \\ k_j > 0}}^N\varphi(z_j) \Bigg|\frac{\rho^{-k_j (n-\ell-1)}\Psi^{(k_j)}_{n-\ell-1}[f](z_j)}{k_j!\varphi(z_j)}
- \langle\eta, f\rangle^{k_j}L_{k_j}(z_j)\Bigg|\Bigg]\Bigg](x).
\label{bound1}
\end{align}
Thanks to the induction hypothesis, we can take $n$ sufficiently large to ensure that the absolute brackets given in the summands are smaller than any given $\delta > 0$. With the help of \eqref{DCT}, assumption \ref{G1} and the fact that $\varphi$ is uniformly bounded above, for $n$ sufficiently large, we can ensure that the right-hand side of \eqref{bound1} is smaller than $\epsilon$.

  To deal with the tail end of the sum in \eqref{tendsto0}, again using the inequality $|\prod a_i - \prod b_i|\leq \sum|a_i-b_i|$, we see that the sum from $n_0+1$ to $n-1$ in \eqref{tendsto0} is bounded above by
  \begin{align}
  \lim_{n \to \infty} k!&\sum_{\ell = n_0+1}^{n-1} \rho^{-k(\ell + 1)}\rho^\ell\rho^{-\ell}\varphi(x)^{-1}\Psi_\ell\Bigg[\tilde{\mathcal{E}}_\cdot \Bigg[\sum_{[k_1, \dots, k_N]_k^{2+}}\sum_{\substack{j = 1 \\ k_j > 0}}^N\varphi(z_j) \Bigg|\frac{\rho^{-k_j (n-\ell - 1)}\Psi^{(k_j)}_{n-\ell - 1}[f](z_j)}{k_j!\varphi(z_j)} \notag\\
&\hspace{10cm} - \langle\eta, f\rangle^{k_j}L_{k_j}(z_j)\Bigg|\Bigg]\Bigg](x). \label{tailbound}
  \end{align}
  Again, using the induction hypothesis, \eqref{DCT} and \ref{G1}, we can also ensure that  there is a global constant $C$ such that the term $\rho^{-\ell}\varphi(x)^{-1}\Psi_\ell [\cdots]$ in \eqref{tailbound}, is bounded by $C$, for all $\ell\geq 1$. Hence taking $n\to\infty$ in \eqref{tailbound}, we see that \eqref{tailbound} is bounded above by
\begin{equation*}
Ck!{\color{black}\lim_{n \to \infty}}\sum_{\ell = n_0+1}^n\rho^{-k(\ell + 1)}\rho^{\ell}
= Ck!\rho^{-k}\sum_{\ell = n_0 + 1}^\infty \rho^{-\ell(k - 1)},
\end{equation*}
which can be made arbitrarily small by again pushing $n_0$ to larger values. The desired conclusion
now follows. \hfill$\square$

\subsection{Subcritical case, $\rho < 1$}

Again referring to \eqref{kmomentcomplex}, we start by noting that scaling by $\varphi(x)^{-1}\rho^{-n}$ and taking limits as $n\to\infty$, the first  term on the right-hand side converges uniformly to  $\langle \eta, f^\ell\rangle$. 

Next define $[k_1,\cdots, k_N]^{(j)}_k$ as the set of non-negative tuples $(k_1,\dots, k_N)$, such that $\sum_{i = 1}^N k_i=k$ with exactly $j$ of the $k_i$ strictly positive. We can write the second term in the definition of $L_k$ as
\begin{align}
&\sum_{\ell = 0}^\infty \rho^{-(\ell+1)} \left\langle \eta, \tilde{\mathcal{E}}_\cdot\Bigg[ \sum_{[k_1,\cdots, k_N]^{2+}_k}{k \choose k_1, \dots, k_N}\rho^{\#\{j: k_j>0\}\ell}\prod_{j = 1}^N \rho^{-\ell}\Psi^{(k_j)}_{\ell}[f](z_j)\Bigg]\right\rangle\notag\\
&=\frac{1}{\rho}\sum_{j = 2}^k\sum_{\ell = 0}^{\infty}\rho^{(j-1)\ell}\langle \eta, H^{(j)}_{\ell}[f] \rangle
\label{Lkconverges}
\end{align}
where 
\[
H^{(j)}_{\ell}[f](x) =  \tilde{\mathcal{E}}_x\Bigg[
 \sum_{[k_1,\cdots, k_N]^{(j)}_k}{k \choose k_1, \dots, k_N}\prod_{j = 1}^N \rho^{-\ell}\Psi^{(k_j)}_{\ell}[f](z_j) 
 \Bigg].
\]
Using our usual arguments of \eqref{DCT} and the induction hypothesis, we can bound $H^{(j)}_{\ell}$ by a global constant. Hence, recalling that $\rho<1$, we see that the left-hand side of \eqref{Lkconverges} is a convergent series. Hence, we can ensure that for $n$ sufficiently large, the residual summation from $n$ to $\infty$ is arbitrarily small. 

Returning to \eqref{kmomentcomplex} and scaling the second term on the right-hand side by $\varphi(x)^{-1}\rho^{-n}$ and reversing the order of the summation, the remarks in the previous paragraph imply that it suffices to show that
\[
\chi_n[f](x): =\frac{1}{\rho} \sum_{j = 2}^k\sum_{\ell = 0}^{n-1}\rho^{(j-1)\ell} |\varphi(x)^{-1}\rho^{-(n-\ell-1)}\Psi_{n-\ell-1}[H^{(j)}_{\ell}[f]](x)
-\langle\eta, H^{(j)}_{\ell}[f]\rangle|
\]
tends to zero, uniformly for $f \in B(U)$ and $x \in U$.

For any fixed $n_0>0$, for all  $n_0<\ell\leq n-1$, the terms $|\varphi(x)^{-1}\rho^{-(n-\ell-1)}\Psi_{n-\ell-1}[H^{(j)}_{\ell}[f]](x)
-\langle\eta, H^{(j)}_{\ell}[f]\rangle|$ are uniformly bounded by a global constant, say $C$, thanks to the induction hypothesis, \ref{G1} and \eqref{DCT}. Hence, taking $n_0$ sufficiently large, the `upper end of the sum' is bounded above by
\[
C\sum_{\ell = n_0+1}^\infty\rho^{(j-1)\ell},
\]
which tends to zero as $n_0$ tends to infinity as $\rho<1$.

On the other hand, for $\ell\leq n_0$, i.e. the  `lower end of the sum', the induction hypothesis, \eqref{DCT} and (H1) yet again ensure that, since we can uniformly bound $H^{(j)}_\ell[f](x)$ over $j=2,\cdots, k$  and  $\ell\leq n_0$, the terms $|\varphi(x)^{-1}\rho^{-(n-\ell-1)}\Psi_{n-\ell-1}[H^{(j)}_{\ell}[f]](x)
-\langle\eta, H^{(j)}_{\ell}[f]\rangle|$
can be made arbitrarily small as $n\to\infty$, uniformly (in $x\in E$ and $ f\in B(U))$. This ensures that the sum over $\ell\leq n_0$ can be controlled as $n \to \infty$. This completes the proof. \hfill$\square$.

\subsection{Conversion of results to NTE setting at criticality}\label{sec:NTE}
The reader may have noticed a slight difference in the limits stated in the main theorems in \cite{NTE-mom} and those stated in the previous section here. For example, the term $\langle  \eta, \tilde{\mathcal{V}}[\varphi]\rangle$ appearing in the limit given in Theorem \ref{thm:crit} above is replaced by $\langle\varphi_0^\dagger, \Sigma_{\mathtt f}\mathcal{V}[\varphi_0]\rangle$ in Theorem 1 of \cite{NTE-mom}, where $\mathcal{V}$ has the same definition as $\tilde{\mathcal{V}}$ albeit with $\tilde{\mathcal{P}}$ replaced by $\mathcal{P}$.  The reason for this is due to the fact that the discrete time fission process is obtained from a continuous time process, as previously discussed. Let us give a few more details on this. 

Recall from \cite{NTE-mom} that in the case of neutron transport, we have $U = D \times \mathbb S_2 \times (E_\mathtt{min}, E_\mathtt{max})$, where $D \subset \mathbb R^3$ is an open, bounded set, $\mathbb S^2$ is the unit sphere in $\mathbb R^3$ and $0 < E_{\mathtt{min}} \le E_{\mathtt{max}} < \infty$.  We also have $x_i^{(n)} = (r_i^{(n)}, \Omega_i^{(n)}, E_i^{(n)})$ for $i = 1, \dots, N_n$, $n \ge 1$.  Also recall from \cite{NTE-mom} the backwards operators $T^\dagger$, $S^\dagger$ and $F^\dagger$
\begin{align*}
T^\dagger[g](r, \Omega, E) &= -\Omega \cdot \nabla g(r, \Omega, E) + \Sigma_{\mathtt{t}}(r, E)g(r, \Omega, E) \\
S^\dagger [g](r, \Omega, E) &= \Sigma_\mathtt{s}(r, E) \int_{E_\mathtt{min}}^{E_\mathtt{min}}\int_{\mathbb S_2} g(r, \Omega', E') f_\mathtt{s}(r, \Omega \to \Omega', E \to E')\d \Omega' \d E',\\
F^\dagger [g](r, \Omega, E) &= \nu_{\mathtt f}(r, E)\Sigma_\mathtt{f}(r, E) \int_{E_\mathtt{min}}^{E_\mathtt{min}}\int_{\mathbb S_2} g(r, \Omega', E') \chi_\mathtt{f}(r, \Omega \to \Omega', E \to E')\d \Omega' \d E' ,
\end{align*}
and their adjoints $T$, $S$ and $F$. Recall further that $\varphi_0$ and $\varphi_0^\dagger$ solve
\begin{equation}\label{eq:back-flux}
(T^\dagger - S^\dagger)[\varphi_0^\dagger] = \frac{1}{k_0}F^\dagger \varphi_0^\dagger \quad \Leftrightarrow \quad -(-T^\dagger + S^\dagger)^{-1}F^\dagger[\varphi_0^\dagger] = k_0\varphi_0^\dagger,
\end{equation}
and
\begin{equation}\label{eq:fwd-flux}
(T - S)[\varphi_0] = \frac{1}{k_0}F \varphi_0 \quad \Leftrightarrow \quad -(-T + S)^{-1}F[\varphi_0] = k_0\varphi_0.
\end{equation}
 Applying $F^\dagger$ to the right-hand equation of \eqref{eq:back-flux} implies that there exists a measure $\eta$, which is absolutely continuous respect to 
 $\d r\d \Omega\d E$, and whose density (also denoted $\eta$) satisfies
 \begin{equation}\label{eq:eta}
-F^\dagger(-T^\dagger + S^\dagger)^{-1}[\eta] = \eta,
\end{equation}
where $\eta := F^\dagger[\varphi_0^\dagger] = Q_{\mathtt f}$.

\bigskip


Noting that $\tilde{\mathcal E}$ is the expectation operator associated with the branching operator that projects one post-fission generation to the next, i.e. it combines transport and fission, we split $\tilde{\mathcal E}$ into the two components by writing, for $g\in B(U)$,
\begin{equation}
\tilde{\mathcal E}_{(r,\Omega, E)}[\langle g, \mathcal{Z}\rangle] = \mathtt P[\Sigma_{\mathtt f}{\mathcal E}_\cdot[\langle g, \mathcal{Z}\rangle]](r,\Omega, E),
\label{tildenotilde}
\end{equation}
 where
\begin{equation}\label{eq:Q}
\mathtt P[f](r, \Omega, E) = \mathbf E_{(r, \Omega, E)}\Bigg[\int_0^\infty {\rm e}^{-\int_0^s \Sigma_{\mathtt f}(R_u, \Omega_u, E_u) {\rm d} u}f(R_s, \Omega_s, E_s) {\rm d} s\Bigg],
\end{equation}
and $(R_t, \Omega_t, E_t)_{t \ge 0}$ denotes the process whose generator is given (informally) by $-T^\dagger + S^\dagger$. Then the operator $\mathtt P$ propagates a particle from $(r, \Omega, E)$ according to the generator $-T^\dagger + S^\dagger$, killed at rate $\Sigma_{\mathtt f}$.

\begin{definition}\label{psiminus}\rm
With this new notation, we see that semigroup associated with the $n$-th generation of {\it pre}-fission particles can be identified by $\Psi_{n}^-[g] := \Psi_{n-1}[\mathtt P[\Sigma_{\mathtt f} g]]$, for $g\in B(U)$.
\end{definition}

Next, let us consider how some of the expressions in Theorems \ref{thm:crit}, \ref{thm:survival}, \ref{thm:Yaglom}, \ref{thm:super} and \ref{thm:sub} translate to the specific form in the setting of neutron transport.

\begin{lemma}  We have
\begin{itemize}
\item[(i)]  $\langle Q_{\mathtt f}, g \rangle = \langle \varphi_0^\dagger, F[g]\rangle$ for  $g\in B(U)$,
\item[(ii)] $ \Psi_n[\tilde{\mathcal{E}}_\cdot[\cdot]] 
= \Psi^{-}_{n+1}[\mathcal{E}_\cdot[\cdot]]$ and
\item[(iii)] $\langle Q_{\mathtt f}, \tilde{\mathcal{E}}_\cdot[\cdot]\rangle = 
k_{0}\langle\varphi_0^\dagger,  \Sigma_{\mathtt f}{\mathcal E}_\cdot [\cdot]\rangle $
\end{itemize}
\end{lemma}

%

\begin{proof}
The first identity is an easy consequence of the fact that  $Q_{\mathtt f} = F^\dagger[\varphi_0^\dagger]$, and the second identity follows immediately from Definition \ref{psiminus}.

For the third identity,  we have from \eqref{tildenotilde},
\[
\quad\langle Q_{\mathtt f}, \tilde{\mathcal{E}}_\cdot[\cdot]\rangle = 
\langle Q_{\mathtt f},  \mathtt P[\Sigma_{\mathtt f}{\mathcal E}_\cdot [\cdot]]\rangle
= 
\langle \mathtt P^\dagger[Q_{\mathtt f}],  \Sigma_{\mathtt f}{\mathcal E}_\cdot [\cdot]\rangle,
\]
where $\mathtt P^\dagger$ is the adjoint of $\mathtt P$. On the associated $L^2$ space, one can associate the operator $\mathtt P$ with the resolvent operator $-(-T^\dagger + S^\dagger)^{-1}$, see \cite{SNTE3}. Thus,   the operator $\mathtt P^\dagger$ can be associated with the resolvent operator $-(T^\dagger + S^\dagger)^{-1}$. Using this with the definition of $Q_{\mathtt f}$, we have
\begin{equation}
  \mathtt P^\dagger[Q_{\mathtt f}] = -(-T^\dagger + S^\dagger)^{-1}[Q_{\mathtt f}] = -(-T^\dagger + S^\dagger)^{-1}F^\dagger[\varphi_0^\dagger] = k_{0}\varphi_0^\dagger,
  \label{Qdagger}
\end{equation}
where the final equality follows from \eqref{eq:back-flux}.  It now follows that 
\[
\langle \mathtt P^\dagger[Q_{\mathtt f}],  \Sigma_{\mathtt f}{\mathcal E}_\cdot [\cdot]\rangle = 
k_{0}\langle \varphi_0^\dagger ,  \Sigma_{\mathtt f}{\mathcal E}_\cdot [\cdot]\rangle,
\]
as required.
\end{proof}

\begin{rem}\rm
We note that part $(iii)$ of the above lemma implies that $\langle  Q_{\mathtt f}, \tilde{\mathcal{V}}[\varphi_0]\rangle = k_{0}\langle \varphi_0^\dagger, \Sigma_{\mathtt{f}} \mathcal{V}[\varphi_0]\rangle$.
\end{rem}


\section*{Appendix}

\renewcommand{\theequation}{A.\arabic{equation}}
\renewcommand{\thelemma}{A.\arabic{lemma}}
\renewcommand{\thetheorem}{A.\arabic{theorem}}
\renewcommand{\thesubsection}{A.\arabic{subsection}}
\renewcommand{\thesection}{A.\arabic{section}}
   
\setcounter{equation}{0}
\setcounter{theorem}{0}
 \setcounter{lemma}{0}
 \setcounter{section}{0}
 \setcounter{subsection}{0}

We need a special type of convergence result that extends the asymptotic stability in \ref{G1}. In order to state it, let us introduce a class of functions $\mathcal{C}$ on 
 $B(U)\times E\times \mathbb{N}\times \mathbb{N}$ such that $F$ 
 belongs to class $\mathcal{C}$ if it has a `limit partner' $\hat{F}$ on $B(U)\times E\times \mathbb{N}\times \mathbb{N}$ such that for all $x\in E$ and $g\in B(U)$, the limit
  \begin{equation}
\check{F}[g] (x): =\lim_{n\to\infty}\frac{1}{n}\sum_{\ell = 0}^{n-1}   \hat{F}[g](x, \ell,n)  
\label{A2}
  \end{equation}
  exists, 
  \begin{equation}
  \lim_{n\to\infty} \sup_{   g\in B(U)}
\Bigg|
\frac{1}{n
}\sum_{\ell = 0}^{n-1} \langle   \eta, \varphi \hat{F}[g](\cdot, \ell, n) \rangle 
-\langle \eta ,\varphi \check{{F}}[g]\rangle 
\Bigg|=0,
\label{triangle}
  \end{equation}
\begin{equation}
\sup_{x \in U, \ell\leq n\in \mathbb{N}, g\in B(U)}| \varphi(x) \hat{F}[g](x, \ell,n)| <\infty,
\label{ass1}
\end{equation}
and 
\begin{equation}
\lim_{n\to\infty} \sup_{x \in U, \ell\leq n , g\in B(U)}\varphi(x)| F[g](x,\ell, n)-\hat{F}[g](x,\ell, n)|  = 0.
\label{ass2}
\end{equation}

\bigskip

\begin{theorem}\label{ergodic}
Assume \ref{G1} holds, $\rho = 1$ and that $F\in\mathcal{C}$. 
Define 
\[
\Xi_n = \sup_{x\in E, g\in B(U)}\Bigg|  \frac{1}{n}
\sum_{\ell = 0}^{n-1} \frac{1}{\varphi(x)}\Psi_\ell[\varphi F[g](\cdot , \ell, n)](x)\d u - 
\langle \eta\varphi , \check{{F}}[g]\rangle
\Bigg|, \qquad t\geq0.
\]
Then
\begin{equation}
\sup_{n\geq 2}\Xi_n<\infty \text{ and }\lim_{n\to\infty}\Xi_n = 0.
\label{erglim}
\end{equation}
\end{theorem}
\begin{proof}
Because of the assumption \eqref{triangle}, the triangle inequality implies that it suffices  to  show  
\[
  \lim_{n\to\infty}\sup_{x\in E, g\in B(U)} \Bigg| 
   \frac{1}{n}
\sum_{\ell = 0}^{n-1}\Big(\frac{1}{\varphi(x)}\Psi_\ell[\varphi F[g](\cdot , \ell, n)](x)  - 
\langle\eta\varphi, \hat{F}[g](\cdot, \ell,n)  \rangle \Big)
\Bigg|= 0.
\]
To that end, first note that,
\begin{align*}
 & \Bigg|  \frac{1}{n}\sum_{\ell = 0}^{n-1}\Big( \frac{1}{\varphi(x)}\Psi_\ell[\varphi F[g](\cdot , \ell, n)](x)  - 
\langle\eta\varphi, \hat{F}[g](\cdot, \ell,n)  \rangle \Big)\Bigg| \notag\\
  &\le   \frac{1}{ \varphi(x)n}\sum_{\ell = 0}^{n-1} \Psi_\ell\Big[\Bigg|\varphi F[g](\cdot , \ell, n) -\varphi \hat{F}[g](\cdot , \ell,n)  \Bigg|\Big](x) \\
  &\qquad + \Bigg| \frac{1}{n}\sum_{\ell = 0}^{n-1}\Big(
  \frac{1}{ \varphi(x)}\Psi_\ell[\varphi \hat{F}[g](\cdot , \ell,n)] -  \langle\eta\varphi, \hat{F}[g](\cdot, \ell,n)  \rangle\Big)\Bigg|.
\end{align*}
Due to assumption \eqref{ass2}, for $n$ sufficiently large, the first term on the right-hand side above can be controlled by 
\[
  \frac{1}{ \varphi(x)n}\sum_{\ell = 0}^{n-1}\Psi_\ell[\varepsilon](x).
\]
Combining this with the above inequality yields
\begin{align}
   &\sup_{x\in E, g\in B(U)}\Bigg|  \frac{1}{n}\sum_{\ell = 0}^{n-1}\Big(\frac{1}{\varphi(x)}\Psi_\ell[\varphi F[g](\cdot , \ell, n)](x)  - 
\langle\eta\varphi, \hat{F}[g](\cdot, \ell,n)  \rangle \Big)\Bigg| \notag\\
   &\hspace{3cm}\le \sup_{x \in U}  
   \Bigg| \frac{1}{ n}\sum_{\ell = 0}^{n-1}\Big( \frac{1}{ \varphi(x)}
    \Psi_\ell[\varepsilon](x) -\langle \varepsilon, \eta\rangle  \Big)\Bigg|   
    + \varepsilon \Vert \eta \Vert_1\notag\\
   &\hspace{4cm}+ \sup_{x\in E, g\in B(U)} \Bigg| \frac{1}{ n}\sum_{\ell = 0}^{n-1}\Big(
  \frac{1}{ \varphi(x)}\Psi_\ell[\varphi \hat{F}[g](\cdot , \ell,n)] -  \langle\eta,\varphi \hat{F}[g](\cdot, \ell,n)  \rangle\Big)\Bigg|.
   \label{twobounds}
\end{align}
Suppose now we choose $\varepsilon>0$ and take  $n_0\in\mathbb{N}$ such that $\Delta_n$, defined in \ref{G1}, is smaller than $\varepsilon$ for all $n \ge n_0$. Note that both summations in \eqref{twobounds} can be split at $n_0$, with the summations from $1$ to $n_0$ disappearing in the limit. This is due to the uniform bound of the summands under the assumptions \eqref{ass1}, \ref{G1} and the scaling of these partial sums by $1/n$. For the remaining sums from $n_0+1$ to $n-1$ on the right-hand side of \eqref{twobounds}, we use the choice of $n_0$, \eqref{ass1} and \ref{G1}, to ensure that the limits as $n\to\infty$ are arbitrarily small. This gives us both of the statements in \eqref{erglim}.
\end{proof}

\section*{Acknowledgments}
This work was partially supported by the EPSRC grant EP/W026899/1. The authors report there are no competing interests to declare.

\bibliographystyle{plain}                                                                           
\bibliography{bibliography}